\documentclass{article}
\usepackage[utf8]{inputenc}
\title{Multi-Agent Distributed and Decentralized Geometric Task Allocation}
\author{Michael Amir, Yigal Koifman, Yakov Bloch\\ Ariel Barel, and Alfred M. Bruckstein}
\usepackage{xcolor}
\definecolor{linkcolor}{rgb}{0.0, 0.0, 0.35}
\usepackage{graphicx}
\usepackage{amsmath}
\usepackage{amsfonts}
\usepackage{amssymb}
\usepackage{amsthm}
\usepackage{enumitem}
\usepackage{breqn}
\usepackage[font=small,labelfont=bf]{caption}

\usepackage{changepage}
\usepackage{upgreek}
\newcommand{\figureWidthAdjustment}{-1.5cm}
\usepackage{multirow}
\usepackage{hhline}
\usepackage{makecell}
\usepackage[T1]{fontenc}
\usepackage[utf8]{inputenc}
\usepackage{xcolor}
\usepackage{array}
\usepackage{subcaption}
\usepackage{hyperref}

\begin{document}

\maketitle

\begin{abstract}
We consider the general problem of geometric task allocation, wherein a large, decentralised swarm of simple mobile agents must detect the locations of tasks in the plane and position themselves nearby. The tasks are represented by an \textit{a priori unknown} demand profile  $\Phi(x,y)$ that determines how many agents are needed in each location. The agents are autonomous, oblivious and indistinguishable, and have finite sensing range. They must configure themselves according to $\Phi$ using only local information about $\Phi$ and about the positions of nearby agents. All agents act  according to the same local sensing-based rule of motion, and cannot explicitly communicate nor share information. 

We propose an optimization-based approach to the problem which results in attraction-repulsion dynamics. Repulsion encourages agents to spread out and explore the region so as to  find the tasks, and attraction causes them to accumulate at task locations. We derive this approach via gradient descent over an appropriate ``error'' functional, and test it extensively through numerical simulations.\\
The figures in this work are snapshots of simulations that can be viewed online at 
\begin {center}
\textcolor{linkcolor}{\url{https://youtu.be/kyUiGYSaaoQ}}
\end {center}
\end{abstract}

\section{Introduction}

This work explores the topic of deploying a robotic swarm of autonomous mobile agents over a region to locate and carry out an \textit{a priori unknown} set of tasks. %Each task requires some number of agents to accumulate in its proximity.
The spatial location of the tasks and the number of agents required to complete them are not given to the agents in advance, and may even change over time. The goal of the agents is to explore the environment to find the tasks, and to position themselves in the region based on the requirements of each task. The agents must also relocate in response to changes in the set of tasks - for example, agents that complete a given task should go on to help other agents complete their tasks.  Examples of this kind of setting include search and rescue missions, where agents must find and assist an unknown number of people, or forest fires, where the spread and intensity of fire evolves over time and requires varying numbers of firefighting drones to cover. %, and pre-emptive allocation, where agents must distribute themselves in space to be able to respond as quickly as possible to a future event (e.g., a forest fire) based on the likelihood it occurs in a given location (as determined by, e.g., smoke and tree density). 

The problems and solutions considered in this work are motivated by common assumptions made in the field of swarm robotics. The objective of swarm robotics is to coordinate a robotic task-force made up of a very large number of simple mobile agents. The agents are assumed to be disposable and redundant: there are more than enough of them to satisfy the demands of all tasks even if some should crash or become lost. Swarm robotics is uniquely positioned to handle task allocation in unknown environments, because the agents can quickly cover a very large area to locate the tasks, and because there are enough agents that we need not worry about some of the agents not finding a task to work on. The main goals of this work are:

\begin{enumerate}
    \item To describe the general problem setting of geometric task allocation for swarms of simple autonomous agents with limited sensing range.
    \item To study several different task allocation problems in this setting.
    \item To make the observation that sensing range limitations can be overcome by equipping a sufficiently large swarm of agents with local  attraction-repulsion dynamics.
\end{enumerate}

\textbf{The model.} Assume $\mathcal{N}$ identical mobile agents are initiated at arbitrary locations within some closed subregion $\mathcal{L}$ of the plane $\mathbb{R}^2$ (such as the unit square $\mathcal{L} = [0,1]\times[0,1]$) and are able to move about $\mathbb{R}^2$ freely. The agents seek to organize themselves within $\mathbb{R}^2$ in a manner determined by a  \textit{demand profile} $\Phi(x,y)$ representing the requirements of tasks. $\Phi(x,y)$ is assumed to be  positive inside $\mathcal{L}$ and $0$ outside of it. Different kinds of demand profiles may be considered: for example, $\Phi(x,y)$ could indicate the required number of agents near position $(x,y)$, or $\Phi(x,y)$ could be a probability density function representing the \textit{proportion} of agents that should be near $(x,y)$ (as in Cortes et al. \cite{cortes2004coverage}), or $\Phi(x,y)$ could be some heat map that needs to be ``covered'' by agents (as in the signal coverage problem we describe below). 

Let us denote the position of the $i$th agent at time $t$ as $\vec{p}_i(t) = (x_i(t),y_i(t))$, and define: \begin{equation}\vec{\textbf{q}}(t) = (x_1(t), y_1(t), x_2(t), y_2(t), \ldots x_{\mathcal{N}}(t), y_{\mathcal{N}}(t))\end{equation} To determine how well the agents satisfy a given demand profile, we define an error function $\Psi(x,y,\vec{\textbf{q}})$ based on $\Phi$, which measures the degree to which the demand $\Phi$ is satisfied at point $(x,y)$ given the agents' current positions  $\vec{\textbf{q}}$. The agents' goal is to move to a position $\vec{\textbf{q}}$ that minimizes the total error over all locations:

\begin{equation}
     \min\limits_{\vec{\textbf{q}}} \iint_{\mathbb{R}^2}\Psi(x,y,\vec{\textbf{q}}) \,\mathrm{d}x\,\mathrm{d}y
\end{equation}

Agents have no common frame of reference, are not aware of distant agents' positions, and do not know the entire demand profile $\Phi$ in advance. Instead, we assume agents possess local sensing capabilities, such that each agent can sense other agents within a finite distance $V_A$ of itself and knows their location relative to itself, and each agent can sense the values of $\Phi$ within distance $V_A$ from itself. %In other words, an agent located at $(x,y)$ can only detect other agents, and read the values of $\Phi$, within a distance $V_A$ from $(x,y)$.%

Time is discrete. At every time step $t = 0, 1, 2, \ldots$, the agents make some small discrete jump in any direction based on what they sense. The distance an agent can move in a single time step is assumed to be bounded by a parameter $\Delta > 0$.

We assume the agents are \textit{oblivious}, meaning they have no memory of actions and computations from previous time steps \cite{barel2019cometogether}, and we assume agents cannot explicitly communicate with each other. Hence, at a given point in time $t$, each agent must determine its next move based only on the information that it currently senses.

Note that while agents can (locally) sense the demand profile $\Phi$, they cannot sense the error function $\Psi$. Whereas $\Phi$ determines the environmental information that agents can sense about their tasks, $\Psi$ determines the agents' overall objective, i.e., the type of task allocation problem they must solve. The main task allocation problems considered in this work are:

\begin{enumerate}
    \item \textbf{Signal coverage}, wherein each agent is surrounded by a ``signal'' (representing, e.g., its effect on nearby points), and agents must minimize the squared error between their combined signal and $\Phi$, hence ``covering'' $\Phi$.
    \item \textbf{Target assignment}, a highly related problem wherein a finite number of discrete targets (e.g., search and rescue tasks) are placed at unknown locations in the region, and each location requires some predefined number of agents in its vicinity.
    %\item \textbf{Voronoi-based coverage control} or ``vector quantization'', a well-studied problem  \cite{cortes2004coverage} wherein the agents seek to approximate some probability density function $\Phi(x,y)$ (e.g., the probability of a forest fire breaking near $(x,y)$) via weighted Voronoi tessellation. \\
\end{enumerate}

%We consider two different ways of defining both $\Phi$ and $\Psi$, called ``models,'' each motivated by problems in the multi-robot and signal processing literature (see ``Related Work'' below):

%\begin{enumerate}
 %   \item The \textbf{influence model}, wherein a finite number of tasks are located across $\mathcal{L}$, each requiring some predefined number of agents in its vicinity.
    
  %  \item The \textbf{vector quantization model}. Interpret $\Phi(x,y)$ as a probability mass function representing the probability that an event requiring an agents' response occurs at $(x,y)$ (such as a forest fire), and let $\textbf{s}$ be a randomly sampled point from $\Phi$.  In this model, we seek to position the agents so as to minimize the squared distance between $\textbf{s}$ and the closest agent to it.
%    \item The \textbf{electrostatic model}, wherein agents exert repelling forces that counteract attracting forces generated by $\Phi(x,y)$, and seek to arrive at an equilibrium.
%\end{enumerate}

For both of these problems, our goal is to find a corresponding error function and a local rule of motion that minimizes the total error. This problem is a basic distributed and decentralized task allocation issue, and similar problems are often discussed in the literature. However, we could not find any prior work that addresses this issue under the combined assumptions of local sensing, decentralized decision making, and no explicit inter-agent communication.

\begin{figure}[ht]
    \begin{adjustwidth}{\figureWidthAdjustment}{\figureWidthAdjustment}
     \centering
     \begin{subfigure}[t]{0.32\columnwidth}\hfil
         \centering
         \includegraphics[width=\linewidth]{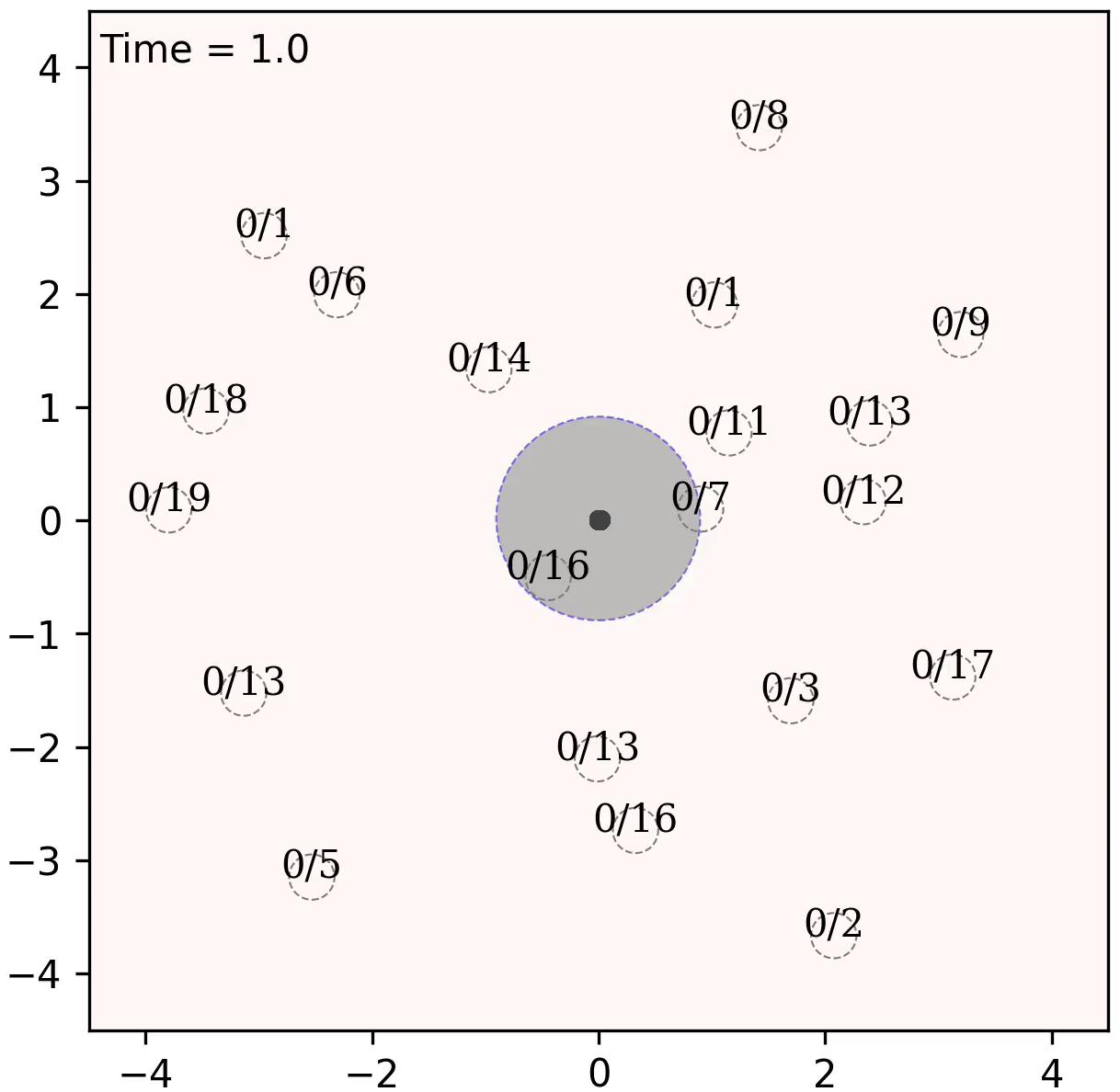}
     \end{subfigure}
     \begin{subfigure}[t]{0.32\columnwidth}\hfil
         \centering
         \includegraphics[width=\linewidth]{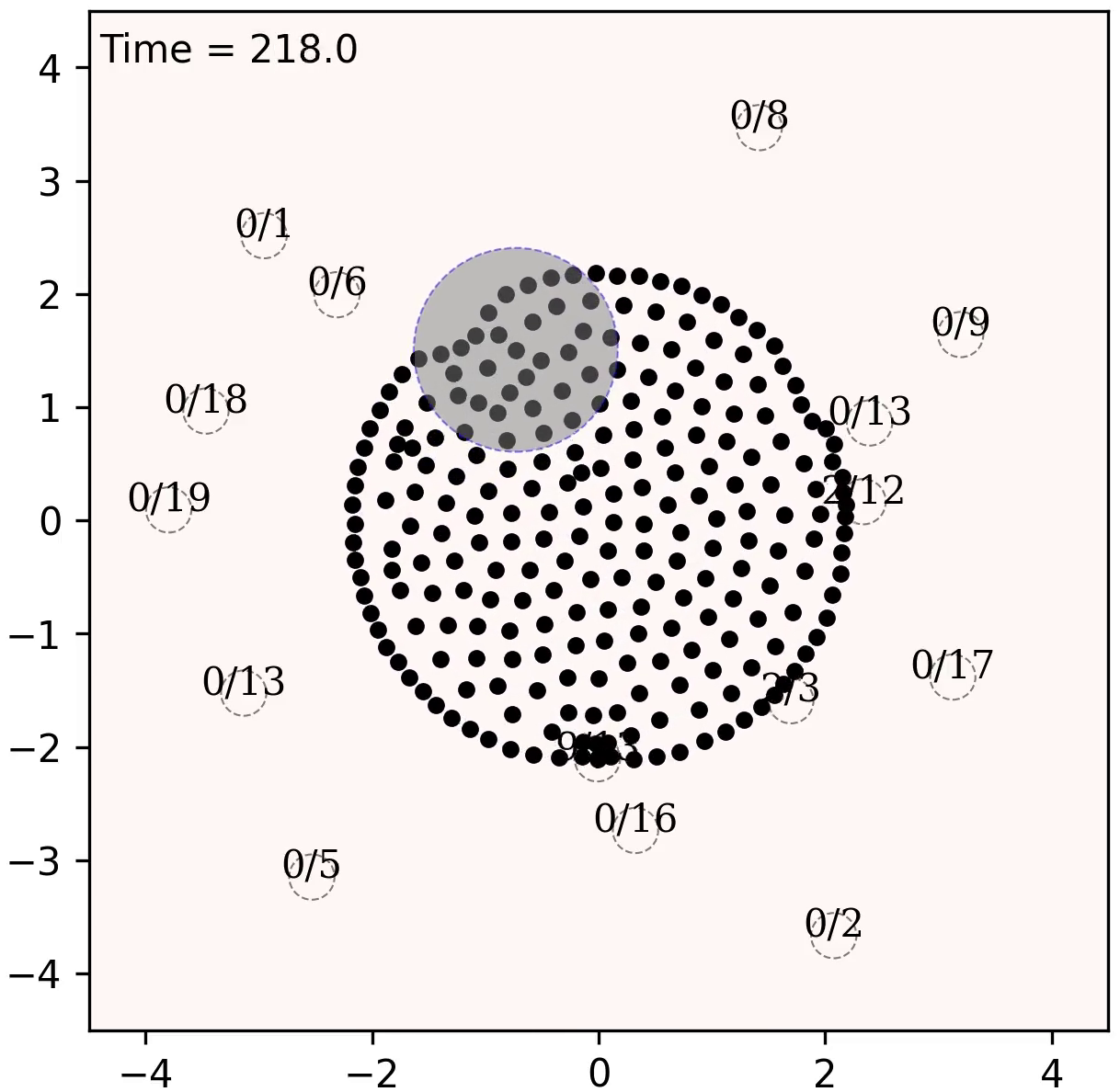}
     \end{subfigure}
     \begin{subfigure}[t]{0.32\columnwidth}\hfil
         \centering
         \includegraphics[width=\linewidth]{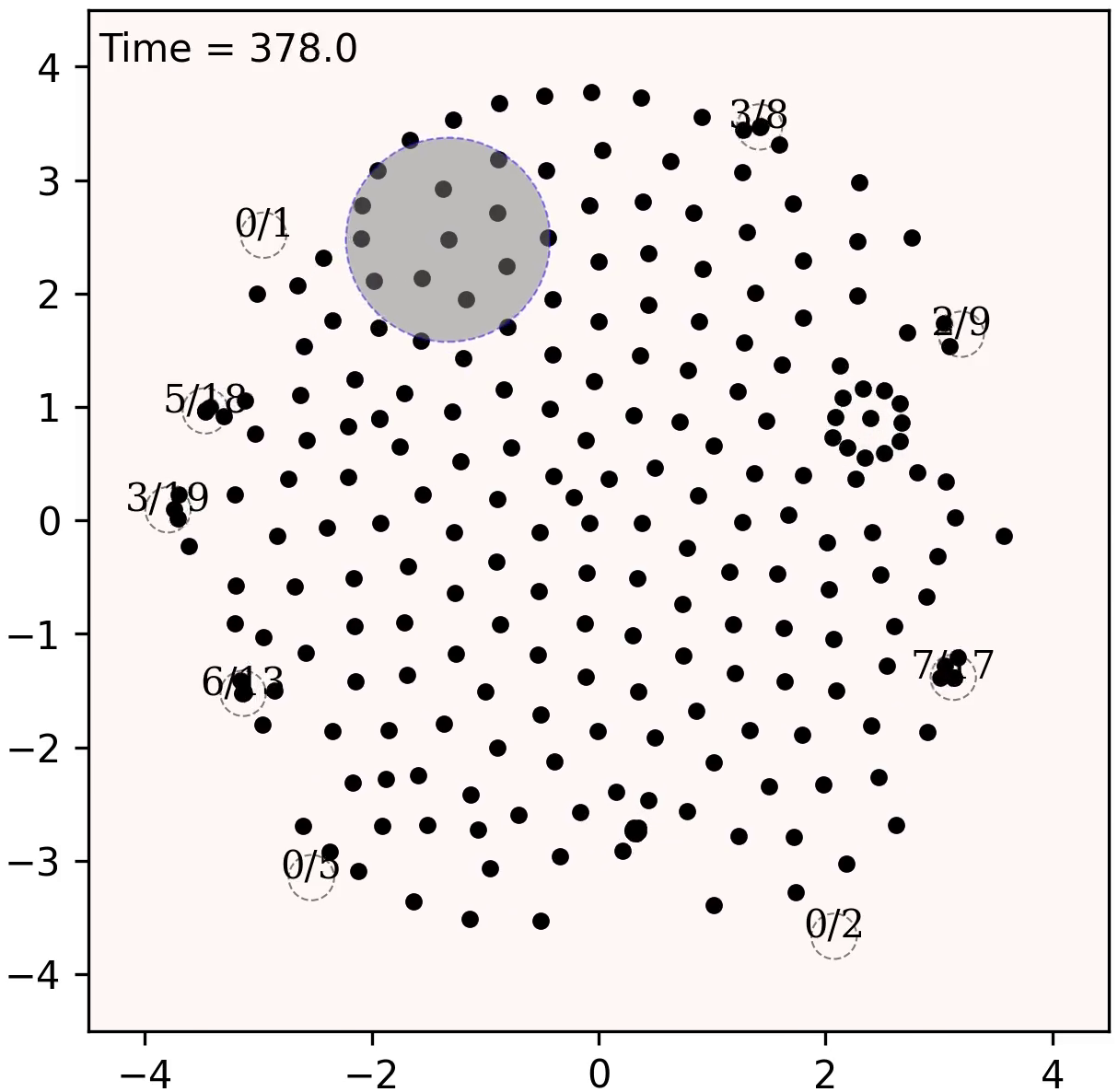}
     \end{subfigure}
     
     \begin{subfigure}[t]{0.32\columnwidth}\hfil
         \centering
         \includegraphics[width=\linewidth]{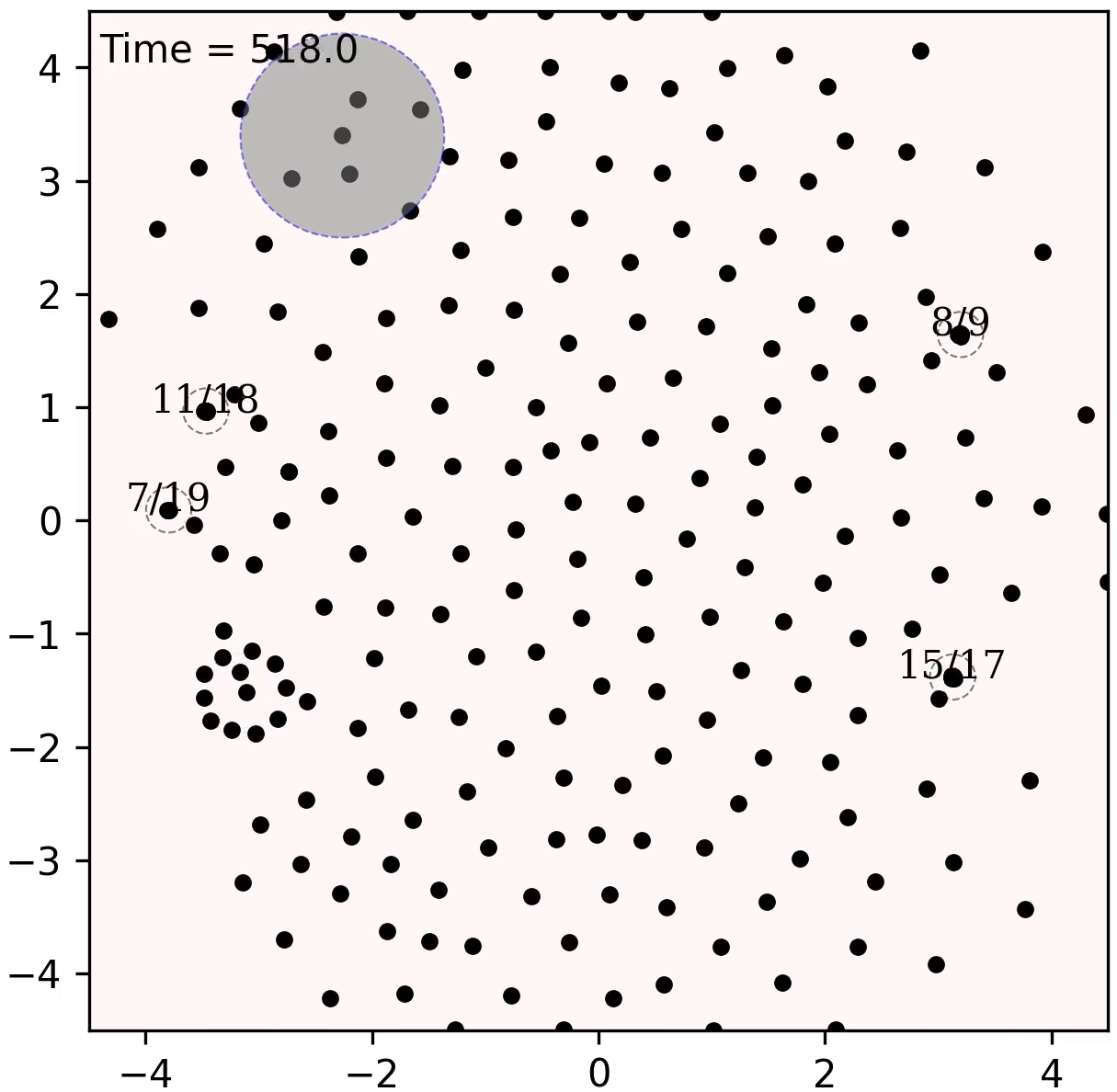}
     \end{subfigure}
     \begin{subfigure}[t]{0.32\columnwidth}\hfil
         \centering
         \includegraphics[width=\linewidth]{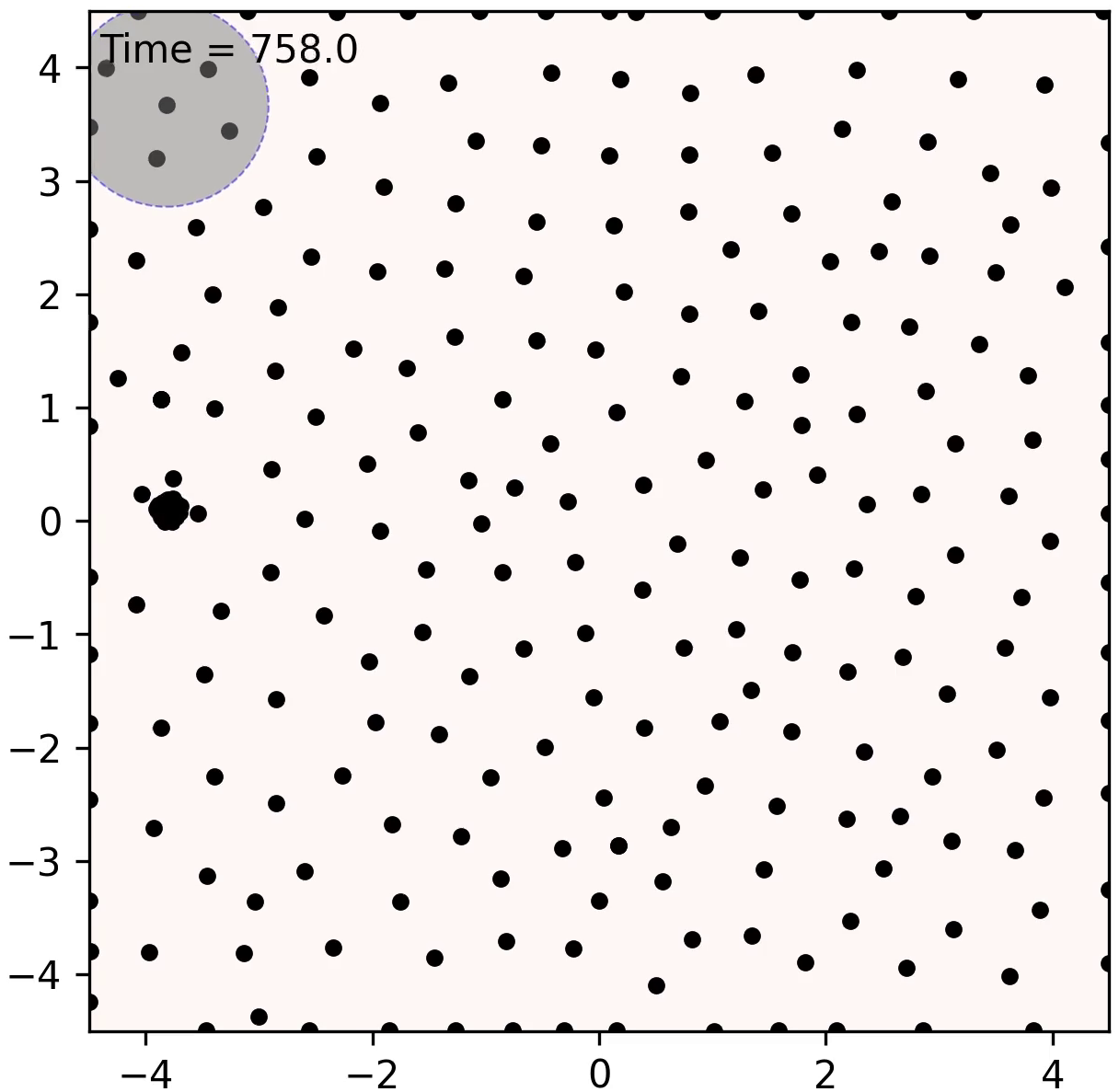}
     \end{subfigure}
     \begin{subfigure}[t]{0.32\columnwidth}
         \centering
         \includegraphics[width=\linewidth]{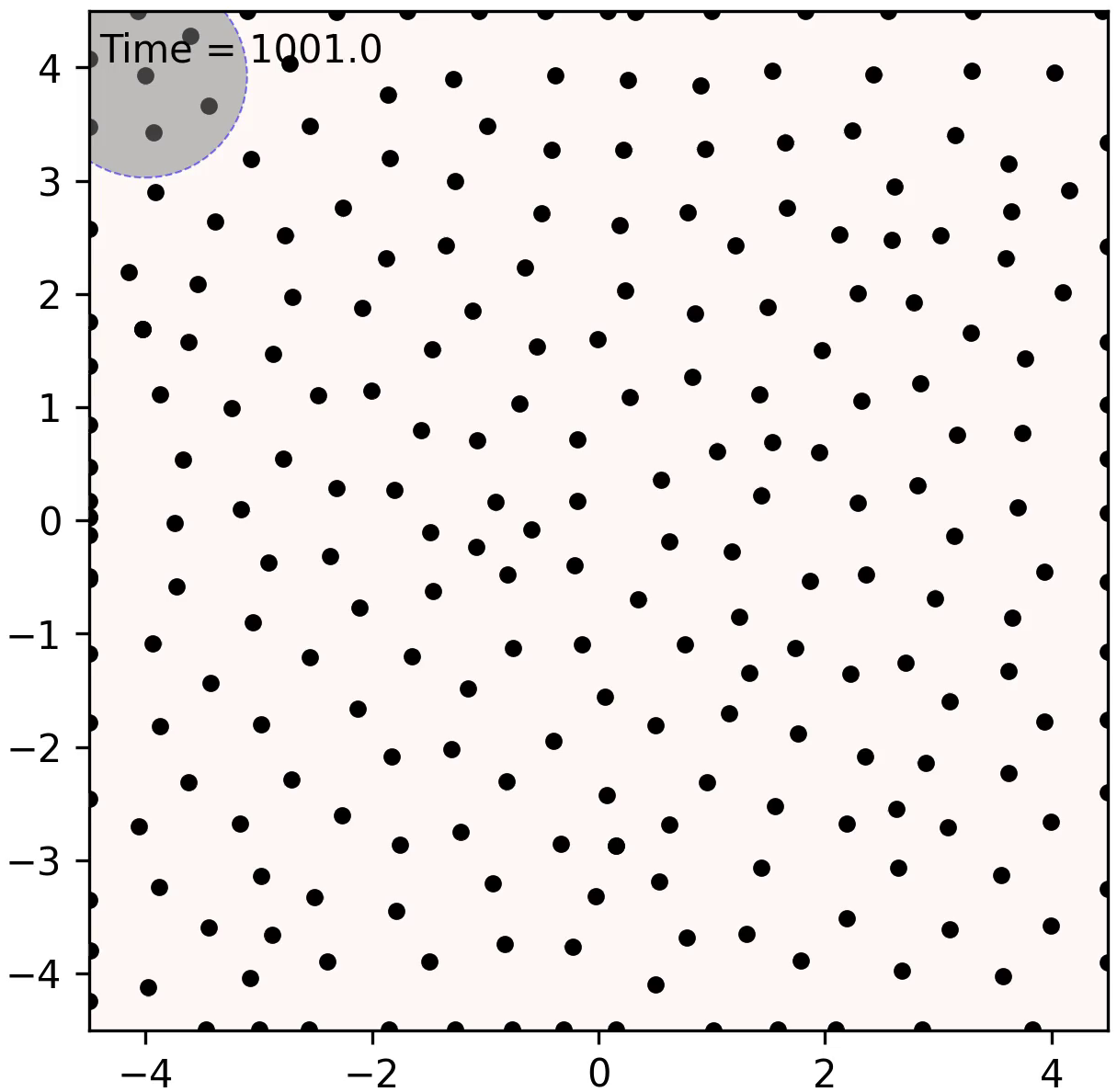}
     \end{subfigure}
     \end{adjustwidth}
     \caption{
     Simulation of a target assignment scenario where targets disappear upon reaching the demanded number of agents. Agents move according to the electrostatic attraction-repulsion dynamics outlined in  (\ref{eq:electrostaticdynamicsinfinitevisibility}). The current time step is written in the top left corner of each frame. Agents' sensing range is depicted by the transparent gray  disk (sensing range is depicted for a single agent to prevent visual clutter). $200$ agents begin at the center of the region and expand outward, finding targets. The initial total demand of the targets is $204$.}
     \label{fig:Sim1_target_destruction}
\end{figure}

Such conservative assumptions about the agents' capabilities result in algorithms that are resilient to sudden changes or crashes. For example, removing agents from the system, or altering the demand profile $\Phi$ in real time (e.g., due to a changing environment or changing set of tasks) does not break any of our algorithms, but instead leads the agents to correctly reconfigure their constellation. The agents can thus seamlessly react to changes in the environment or the tasks. As an example, Figure \ref{fig:Sim1_target_destruction} shows a target assignment scenario where agents, starting in the middle of the region, search for tasks placed randomly within the region via attraction-repulsion dynamics. Each task demands a different number of agents to complete. When a sufficient number of agents reach a task, the task is completed and removed from $\Phi$, freeing agents to move to other tasks.

%%%%%%%%%%%%%%%%%%%%%%%%%%%%%%%%%%%%%%%%%%%%%%%%%%%%%%%%%%%%%%%%%%%%%%%%
\section{Related Work}

Although the literature on swarm task allocation and task-based positioning is varied and deep, covering centralized \cite{richards2002coordination,liu2012large}, decentralized  \cite{parker2002distributed,bethke2008uav,michael2008distributed,alighanbari2007robust,jevtic2011distributed}, and non-geometric \cite{shehory1995task,luo2014provably,pham2018decentralized} models, we could not find any prior work that combined all our assumptions of local sensing, decentralized, oblivious decision making, and no explicit communication. These prior works thus take advantage of assumptions and methods that are not available to us, such as centralized trajectory planning, prior knowledge of the world, or direct inter-agent communication. %These assumptions do appear \textit{separately} in many works: e.g., \cite{schwager2009decentralizedlimitedvisibility} talks about decentralized coverage control where each robot has only local information, but the agents are not oblivious. 

The closest works to ours that we could find  are works in dynamical systems along the lines of Gazi et al. \cite{gazipasino2004class,gazipasino2004stability}, Cortes et al. \cite{cortes2004coverage} and Schwager et al.  \cite{slotineschwager2011unifying}. Our work differs from these because of our sensing range constraints, because of the different kinds of task allocation problems we explore, such as signal coverage, and in several theoretical aspects which we outline here. In works by Gazi et al. \cite{gazipasino2004class,gazipasino2004stability}, as well as related works such as \cite{weixing2006novelAR1,bo2005swarmAR2}, various types of attraction-repulsion functions for stable swarm aggregation are considered, occasionally alongside a demand profile-esque ``nutrient profile'' whose gradient guides the agents' dynamics. In contrast to Gazi et al. we do not assume attraction-repulsion dynamics but show how they arise from minimizing the square error of a difference of signal functions. In \cite{schwager2009decentralizedlimitedvisibility,cortes2004coverage} demand profile functions similar to our $\Phi$ are used to guide decentralised agents toward a desirable density distribution and deployment formation, for example based on Voronoi diagrams. \cite{slotineschwager2011unifying} and \cite{cortes2004coverage}  represent environmental data as a multiplicative weight that measures the global importance of a given point in the region with regards to the agents' objective. In contrast, our work treats the environment, $\Phi$, as a separable feature of the agents' sensing data on which the agents perform local computations. We believe that this approach is more applicable in limited-visibility settings, where agents might not have access to sufficient information to gauge the global importance of a point in the region.

%Our work differs from these because of our sensing range constraints, and due to the kinds of task allocation problems we explore, such as signal coverage. Besides these differences, we note that works such as \cite{slotineschwager2011unifying} and \cite{cortes2004coverage} prefer to represent environmental data as a multiplicative weight that measures the global importance of a given point in the region with regards to the agents' objective. In contrast, our work treats the environment, $\Phi$, as a separable feature of the agents' sensing data on which the agents perform local computations. We believe that this approach is more applicable in limited-visibility settings, where agents might not have access to sufficient information to gauge the global importance of a point in the region.

%In the last section of this work we note that the  Voronoi tessellation dynamics of \cite{cortes2004coverage} can be described in terms of our model, and experiment with a limited-visibility variant of it which is motivated by the assumption that there is a large number of redundant agents in the swarm.

Mathematically, the motion dynamics explored in this work can be related to well-studied dynamical systems and techniques in multi-agent systems, signal processing and physics. We outline these prior works here, while also pointing out crucial differences between their settings and ours.

Halftoning is a graphical technique that attempts to simulate a continuous image using a finite set of points  \cite{gilliatt1983study}. The image is represented by a 2D function $\Phi(x,y)$. In electrostatic halftoning, the points are treated as particles that repel each other, whereas $\Phi(x,y)$ determines the local magnitudes of a field of attracting forces \cite{gwosdek2014fast,schmaltz2010electrostatic,teuber2011dithering}. The sum of these repelling and attracting forces causes the particles to position themselves so as to approximate the image expressed by $\Phi$ - a task-allocation-esque problem. Such models assume that an agent can feel the forces exerted by $\Phi$ and other agents globally (albeit to a degree inversely proportional to the agents' distance from the force's origin), whereas we assume that agents are not aware of what happens beyond their visibility range $V_A$. However, it turns out that given a large enough number of agents, halftoning techniques can serve as effective task allocation algorithms even under such constraints--we explore this topic in Section \ref{section:electrostatictargetassignment}.

%Cortes et al.'s deployment algorithm \cite{cortes2004coverage}, which we also mention in this work, can be related to 

%Vector quantization is a technique in signal processing that enables the modelling of a probability mass function via a finite set of points \cite{gray1984vectorquantization}. This concept can be related to 

%was the inspiration for the ``vector quantization model'' of this work, wherein we treat mobile agents as vectors and $\Phi$ as a probability density function that they must approximate. One of the algorithms we propose is an adaptation of Lloyd's quantization algorithm  (see, e.g., \cite{max1960quantizing,lloyd1982least,du1999centroidallloyds}) to a multi-agent setting. A similar effort was undertaken in \cite{cortes2004coverage}, but their algorithm is different from ours and cannot work under the assumption of a constant, predefined sensing range.

In addition to works such as  \cite{cortes2004coverage,slotineschwager2011unifying},  the concept of using a probability mass function to dictate the desirable density of agents at a given location has also been explored in such contexts as ``Optimotaxis,'' Markov chain-based task allocation, and convex optimization \cite{mesquita2008optimotaxis, demir2014density1, density3}. However, unlike our present setting, these prior works assume either communication or global knowledge about the desired density $\Phi(x,y)$. 

When $\Phi(x,y)$ is a constant function, our agents will spread uniformly inside $\mathcal{L}$, thus achieving \textit{uniform dispersion}. Various works have been written on the uniform dispersion of agents in unknown regions and, more broadly, the coverage of unknown regions via a small number of agents. Several of these works share our assumptions of limited visibility and no communication \cite{uniformdispersion,hsianguniform,michaeluniform,barrameda2013uniform,hideg2017uniform}, although they often assume non-oblivious agents. Moreover, some works in this domain have used potential fields \cite{reif1999potentialfields,howard2002mobilepotentialfields}. However, these works are not directly comparable to ours; the goal of coverage and/or uniform dispersion is rather different from the task of positioning agents according to a non-uniform demand profile.

The general field of non-communicating, oblivious, visibility-constrained agents is sometimes referred to as \textit{ant-like swarm robotics}; we refer the readers to the surveys \cite{barel2019cometogether,altshuler2018swarms}. For further reading on task allocation in multi-agent systems, we refer the readers to the surveys \cite{tang2010survey,amador2014dynamictaskallocationsurvey}.

%%%%%%%%%%%%%%%%%%%%%%%%%%%%%%%%%%%%%%%%%%%%%%%%%%%%%%%%%%%%%%%%%%%%%%%%
%\section{The Preamble}

%You will be assigned a submission number when you register the abstract 
%of your paper on \textit{EasyChair}. Include this number in your 
%document using the `\verb|\acmSubmissionID|' command.

%Then use the familiar commands to specify the title and authors of your
%paper in the preamble of the document. The title should be appropriately 
%capitalised (meaning that every `important' word in the title should 
%start with a capital letter). For the final version of your paper, make 
%sure to specify the affiliation and email address of each author using 
%the appropriate commands. Specify an affiliation and email address 
%separately for each author, even if two authors share the same 
%affiliation. You can specify more than one affiliation for an author by 
%using a separate `\verb|\affiliation|' command for each affiliation.

%Provide a short abstract using the `\texttt{abstract}' environment.
 
%Finally, specify a small number of keywords characterising your work, 
%using the `\verb|\keywords|' command. 

%%%%%%%%%%%%%%%%%%%%%%%%%%%%%%%%%%%%%%%%%%%%%%%%%%%%%%%%%%%%%%%%%%%%%%%%
\section{Multi-agent task allocation}
\label{section:multiagenttaskallocation}

In this section we consider two task allocation problems for limited-visibility mobile agents. The first problem we consider is \textit{signal coverage}, in which a swarm of mobile agents surrounded by signals must organize in the region so that their combined signal approximates an \textit{a priori unknown} demand profile $\Phi$. The second, related problem is \textit{target assignment}, in which there are $\mathcal{K}$ targets at a priori unknown locations, and we want to bring a certain number of agents to the location of each target.

We approach these problems through gradient descent on an appropriate error function. We show that this gradient descent leads to \textit{attraction-repulsion dynamics}, wherein agents repulse each other and are attracted to locations in the region determined by $\Phi$. These dynamics' effectiveness can be explained as follows: repulsive forces agents affect each other with cause the swarm to expand uniformly, thus covering the region of interest $\mathcal{L}$ and discovering the demand profile $\Phi$. Attractive forces, on the other hand, cause agents to accumulate according to tasks' requirements (e.g., in target assignment, we want agents to accumulate at the locations of targets, so we have these locations exert attractive forces). The number of agents needed to cover $\mathcal{L}$ depends on the sensing range $V_A$. When the disk of radius $V_A$ is small compared to $\mathcal{L}$, a large number of agents is needed to execute this strategy, making it precisely suited for swarms.

Section \ref{section:signalcoveragemodel} formally describes the signal coverage problem and the corresponding error function $\Psi$. We show that gradient descent over the total error $\mathcal{G}(\vec{\textbf{q}}) = \iint_{\mathbb{R}^2}\Psi(x,y,\vec{\textbf{q}}) \,\mathrm{d}x\,\mathrm{d}y$ leads to an attraction-repulsion-based algorithm for signal coverage.  One approach to the target assignment problem is representing it as a signal coverage problem where there are highly concentrated signals at the location of each target. Hence, Section \ref{section:signalcoveragemodel} can also be applied to the target assignment problem.

A different attraction-repulsion-based approach to target assignment can be derived from Coulomb's law, by treating targets and agents as electrically charged particles. As we shall see, it is not clear whether this solution can \textit{formally} be recovered as a special case of the attraction-repulsion dynamics we derive for signal coverage in Section \ref{section:signalcoveragemodel}, although it is of a very similar form. Motivated by these similarities in form, in Section \ref{section:generalizedattractionrepulsion} we relate both approaches to a general form of attraction-repulsion dynamics.  In this section we also discuss the problem of \textit{scalar field coverage} as a side application of our approach.

\subsection{Signal coverage and target assignment}
\label{section:signalcoveragemodel}

In the \textit{signal coverage problem}, each agent's position is surrounded by a signal $\tilde{d}(x,y)$ which represents a weighted radius of effect (e.g., the effectiveness of a foam cannon  stationed at the agent's position on a fire at $(x,y)$, or the agent's ability to gather data about coordinate $(x,y)$). The agents seek to position themselves in $\mathcal{L}$ so as to minimize the squared difference between their combined signal and an ``environmental signal'' $\Phi(x,y)$ (representing, e.g., the heat map of a forest fire, or the importance of collecting data at coordinate $(x,y)$). Formally, given $\Phi$ and $\tilde{d}$, we define the error function

\begin{equation}
    \Psi(x,y,\vec{\textbf{q}}) = \big(\Phi(x,y) - \sum_{\substack{i = 1}}^{\mathcal{N}} \tilde{d} (x-x_i,y-y_i) \big)^2
\end{equation}

and so our agents' goal is to minimize:

\begin{equation}
     \mathcal{G}(\vec{\textbf{q}}) = \iint_{\mathbb{R}^2}\big(\Phi(x,y) - \sum_{\substack{i = 1}}^{\mathcal{N}} \tilde{d} (x-x_i,y-y_i) \big)^2 \,\mathrm{d}x\,\mathrm{d}y
     \label{eq:psiminimizeinfluence}
\end{equation}

Note that this optimization goal is different from that considered by Cortes et al. \cite{cortes2004coverage}. In \cite{cortes2004coverage}, $\Phi$ is treated as a probability density function, and agents seek to minimize the mean squared error of a randomly sampled point from $\Phi$ to the closest agent. This leads to dynamics wherein agents head toward the center of their respective cell in a (possibly weighted) Voronoi tessellation. In contrast, we minimize the squared error of sums of symmetric functions (representing the signals of agents and targets respectively), which leads to attraction-repulsion dynamics and different optimal configurations from \cite{cortes2004coverage}. 

Intuitively, the signal coverage model is applicable when agents' signals can usefully be combined. For example, several noisy measurements of $(x,y)$ by different drones can be combined to cancel out noise; several foam cannons at agents' positions can be combined to better affect a fire at point $(x,y)$; several agents may be required to complete a task at $(x,y)$. 

\begin{figure}[ht]
  \centering
    \includegraphics[width=55mm]{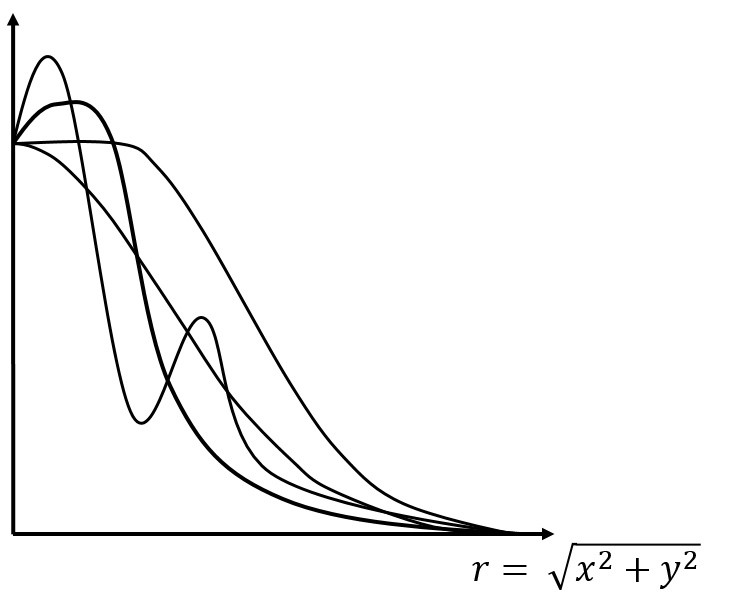}
    \caption{Some possible choices of the signal function $f(r)$. Modifying $f$ will alter the optimal agent signal coverage formation.}
    \label{fig:f_k function}
\end{figure}

$\tilde{d}(x,y)$ is assumed to be a symmetric, almost everywhere differentiable function determined only by $r = \sqrt{x^2 + y^2}$. In other words, there is a signal function $f(r)$ such that $\tilde{d}(x,y) = f(\sqrt{x^2+y^2})$. For some parameter $V$, we assume that 
%$\tilde{d}(0) = 1$ and 
$\tilde{d}(r) = 0$ for all $r > V$ (see Figure \ref{fig:f_k function}). 

%As mentioned in the introduction, we assume the demand profile $\Phi$ is positive everywhere inside  $\mathcal{L}$ and otherwise equals $0$. This bounds the region of interest where the agents search for signals.
 
For the time being, we restrict ourselves to the case where $\Phi(x,y)$ is a weighted sum of $\mathcal{K}$ signals of the same type as the agents' (in Section \ref{section:freddygeneralchoicesofphi}, we extend our approach to more general choices of $\Phi$). The center of the $k$th signal is denoted as $\vec{c}_k = (x_k^S,y_k^S)$. Specifically, we define $\Phi(x,y) = 0$ for all $(x,y) \notin \mathcal{L}$, and otherwise define

\begin{equation}
    \Phi(x,y)= 1 +  \sum_{k = 1}^{\mathcal{K}} n_k \tilde{d} (x-x_k^S,y-y_k^S)
\end{equation}

where $n_k$ is called ``the demand of the $k$th signal center.'' For simplicity, we assume that for all $k$, the disk of radius $V$ centred at $\vec{c}_k$ is completely contained in $\mathcal{L}$. 

The goal of this section is to show signal coverage can be attained through discrete attraction-repulsion dynamics, in which at every time step the $i$th agent changes its position according to:

\begin{equation}
\begin{split}
   \vec{p}_i(t+1) & = \vec{p}_i(t) - \delta  \frac{\vec{v}_i(\vec{\textbf{q}}(t))}{\lVert\vec{v}_i(\vec{\textbf{q}}(t))\lVert}, \textrm{where} \\
    \vec{v}_i(\vec{\textbf{q}}(t)) & = \sum_{\substack{j = 1}}^{\mathcal{N}} F(\lVert \vec{p}_i - \vec{p}_j \rVert)\overrightarrow{p_i p_j} -\sum_{\substack{k = 1}}^{K} n_k F(\lVert \vec{p}_i - \vec{c}_k \rVert) \overrightarrow{p_i c_k}
\end{split} 
\label{eq:attractionrepulsiondynamicsgoal}
\end{equation}

Here, $F(\cdot)$ is some real-valued function derived from $f$ and $\overrightarrow{uv}$ denotes the unit vector from $u$ to $v$. Under these dynamics, every agent is attracted to the centers of the signals, but repulsed by other agents, by an amount that scales according to distance. Since our agents' sensing range is $V_A$, we will be interested in dynamics where $F(r) = 0$ for all $r > V_A$, since then each agent can always compute (\ref{eq:attractionrepulsiondynamicsgoal}).%\footnote{Note that to compute  (\ref{eq:attractionrepulsiondynamicsgoal}), agents must be able to infer the location of the signal center $\vec{c}_k$ when in range of it. One way to formally achieve this is to give $f(r)$ a sharp discontinuity at $r=0$, which makes it so that each signal center is a discontinuous part of the gradient of $\Phi$, but does not affect the optimal formations.}

\textbf{Target assignment.} In the target assignment problem we assume there are $K$ targets at locations $\vec{c}_1, \ldots \vec{c}_K \in \mathcal{L}$, and for $k$, $1 \leq k \leq K$, we want to bring $n_k \geq 1$ agents to the location $\vec{c}_k$. The targets' locations are not known to the agents in advance, and a target can only be detected by an agent at distance $V_A$ or less. As we shall see, the degree to which $\tilde{d}(x,y)$ is concentrated near $(0,0)$ determines the diffusion of the agents around the centers of the signals. For example, when $\tilde{d}$ is heavily concentrated near $(0,0)$, we expect that in an optimal configuration there will be $n_k$ agents very close to the center of the $k$th signal. Through this observation, signal coverage can be applied to the target assignment problem (see Figure \ref{fig:Sim6_signal_coverage_target_assignment}).

In Section \ref{section:electrostatictargetassignment} we describe a different approach to target assignment, based on Coulomb's law.

\subsubsection{Gradient descent and attraction-repulsion dynamics}

To motivate (\ref{eq:attractionrepulsiondynamicsgoal}) as an algorithm for signal coverage, we consider a gradient descent-based rule of motion that enables the agents to minimize $ \mathcal{G}(\vec{\textbf{q}})$ while obeying the restrictions of our model. Our objective is to show that the descent dynamics can be rewritten as attraction-repulsion dynamics of the form (\ref{eq:attractionrepulsiondynamicsgoal}).

In gradient descent, the goal of each agent is to descend the gradient of $\mathcal{G}$ by a small amount. The continuous-time gradient descent dynamics for the $i$th agent are:

\begin{equation}
\def\C{
\begin{bmatrix}
 \frac{\partial \mathcal{G}}{\partial x_i}   \\
\frac{\partial \mathcal{G}}{\partial y_i}  \\

\end{bmatrix}} 
    \frac{d\vec{p}_i}{dt} = -\vec{v}_i(\vec{\textbf{q}}(t)), \textrm{where } \vec{v}_i =  \C
\end{equation}

Since we are working in discrete time dynamics, we may discretize this expression as:
    
\begin{equation}
   \vec{p}_i(t+1) = \vec{p}_i(t) - \delta  \frac{\vec{v}_i(\vec{\textbf{q}}(t))}{\lVert\vec{v}_i(\vec{\textbf{q}}(t))\lVert}
   \label{eq:freddydynamics}
\end{equation}

where $0 < \delta  \leq \Delta$ is some predefined constant. Hence, at every time step, the agents need to compute $\frac{\partial \mathcal{G}}{\partial x_i}$ and  $\frac{\partial \mathcal{G}}{\partial y_i}$. Let us denote $\frac{\partial \tilde{d}(x-x_i,y-y_i)}{\partial x_i}    = \tilde{d}_{x_i}$ and $\frac{\partial \tilde{d}(x-x_i,y-y_i)}{\partial y_i}    = \tilde{d}_{y_i}$. We have that:

%\begin{equation}
%\begin{split}
%    \frac{\partial \mathcal{G}}{\partial x_i} 
%    & =  \frac{\partial }{\partial x_i} \iint_{\mathbb{R}^2}(\Phi(x,y)-\sum_{j=1}^{N} 
%    \tilde{d}(x-x_j,y-y_j))^2 \\
%    & =  2\iint_{\mathbb{R}^2}(\Phi(x,y)-\sum_{j=1}^{N} 
%    \tilde{d}(x-x_j,y-y_j))\tilde{d}_{x_i}(x,y)\,\mathrm{d}x\,\mathrm{d}y \\
%\end{split}
%\label{eq:freddyintegralhalfwaystepx}
%\end{equation}

%And by symmetry: 
\def\D{
\begin{bmatrix}
\tilde{d}_{x_i}(x,y)    \\
\tilde{d}_{y_i}(x,y)  \\
\end{bmatrix}}

\begin{equation}
\def\C{
\begin{bmatrix}
 \frac{\partial \mathcal{G(\vec{\textbf{q}})}}{\partial x_i}   \\
\frac{\partial \mathcal{G(\vec{\textbf{q}})}}{\partial y_i}  \\
\end{bmatrix}}
\C = 2\iint_{\mathbb{R}^2}(\Phi(x,y)-\sum_{j=1}^{\mathcal{N}} 
    \tilde{d}(x-x_j,y-y_j))\D\,\mathrm{d}x\,\mathrm{d}y 
\label{eq:freddyintegralhalfwaystep}
\end{equation}

Recalling that $ \Phi(x,y)= 1 +  \sum_{k = 1}^{\mathcal{K}} n_k \tilde{d} (x-x_k^S,y-y_k^S)$, we split (\ref{eq:freddyintegralhalfwaystep}) into three summands:

\begin{enumerate}[label=(\alph*)]
\item $2\iint_{\mathbb{R}^2}\D\,\mathrm{d}x\,\mathrm{d}y$
\item $ 2\sum_{\substack{k = 1}}^{K}\iint_{\mathbb{R}^2}n_k \tilde{d} (x-x_k^S,y-y_k^S)\D\,\mathrm{d}x\,\mathrm{d}y$
\item $-2\iint_{\mathbb{R}^2}\big(\sum_{j=1}^{\mathcal{N}} 
    \tilde{d}(x-x_j,y-y_j)\big)\D\,\mathrm{d}x\,\mathrm{d}y \\$
\end{enumerate}

To compute (a), we first derive $\tilde{d}_{x_i}(x,y)$:

\begin{equation}
\begin{split}
    \tilde{d}_{x_i}(x,y)
    & = \frac{\partial }{\partial x_i}f(\sqrt{(x-x_i)^2+(y-y_i)^2}) \\
    & = -\dot{f}(\sqrt{(x-x_i)^2+(y-y_i)^2})\frac{x-x_i}{\sqrt{(x-x_i)^2+(y-y_i)^2}}
\end{split}
\end{equation}

We infer that $\tilde{d}_{x_i}(x-x_i,y) = -\tilde{d}_{x_i}(x_i-x,y)$, and analogously  $\tilde{d}_{y_i}(x,y-y_i) = -\tilde{d}_{x_i}(x,y_i-y)$. Hence (a) equals $\begin{bmatrix}0 \\ 0\end{bmatrix}$.

% we first apply a linear translation and rotation $\textbf{\mathrm{M}}$ to the plane

To compute the inner integral $\iint_{\mathbb{R}^2}\tilde{d} (x-x_k^S,y-y_k^S)\tilde{d}_{x_i}(x,y)\,\mathrm{d}x\,y$ in (b), we may assume a rotated frame of reference where $y_i = y_k^S$, and later return to the original frame of reference via the inverse rotation $\boldsymbol{\mathrm{M}}_k^{-1}$ .
We further apply the coordinate transforms $\overline{x} = x - x_i$, $\overline{y} = y - y_i$ to get:%, $(\overline{x},\overline{y}) = (r\cos \theta, r\sin \theta)$ to get:

\begin{equation}
\begin{split}
    &\iint_{\mathbb{R}^2}\ n_k \tilde{d}(x-x_k^S,y-y_k^S)\tilde{d}_{x_i}(x,y)\,\mathrm{d}x\,\mathrm{d}y \\
    & = n_k \iint_{\mathbb{R}^2}\tilde{d} (\overline{x}-\lVert \vec{p}_i - \vec{c}_k \rVert,\overline{y}) \tilde{d}_{x_i}(\overline{x}+x_i,\overline{y}+y_i)\,\mathrm{d}\overline{x}\,\mathrm{d}\overline{y} \\
    & = -n_k \iint_{\mathbb{R}^2}f \big(\sqrt{(\overline{x}-\lVert \vec{p}_i - \vec{c}_k \rVert)^2 + \overline{y}^2}\big)\dot{f}(\sqrt{\overline{x}^2+\overline{y}^2})\frac{\overline{x}}{\sqrt{\overline{x}^2+\overline{y}^2}}\,\mathrm{d}\overline{x}\,\mathrm{d}\overline{y} \\
    %&\hspace*{-45pt} = n_k \iint_{\mathbb{R}^2}\tilde{d} (\overline{x}-\lVert \vec{p}_i - \vec{c}_k \rVert,\overline{y}) \tilde{d}_{x_i}(\overline{x}+x_i,\overline{y}+y_i)\,\mathrm{d}\overline{x}\,\mathrm{d}\overline{y} \\
    %&\hspace*{-45pt} = -n_k \iint_{\mathbb{R}^2}f \big(\sqrt{(\overline{x}-\lVert \vec{p}_i - \vec{c}_k \rVert)^2 + \overline{y}^2}\big)\dot{f}(\sqrt{\overline{x}^2+\overline{y}^2})\frac{\overline{x}}{\sqrt{\overline{x}^2+\overline{y}^2}}\,\mathrm{d}\overline{x}\,\mathrm{d}\overline{y} \\
    %&\hspace*{-45pt} = -n_k \int_{\theta=0}^{2\pi} \int_{r=0}^{\infty} f(\sqrt{(r\cos \theta - \lVert \vec{p}_i - \vec{c}_k \rVert)^2 + (r \sin \theta)^2}) \dot{f}(r) \cos (\theta) r \,\mathrm{d}r\,\mathrm{d}\theta  \\
\end{split}
\label{eq:freddycomputeH}
\end{equation}

We can therefore define a function $F(\cdot)$ such that $(\ref{eq:freddycomputeH}) \triangleq -n_k F(\lVert \vec{p}_i - \vec{c}_k \rVert)$. We have that $F(\lVert \vec{p}_i - \vec{c}_k \rVert) = 0$ when $\lVert \vec{p}_i - \vec{c}_k \rVert \geq 2V$ (since in such cases $\tilde{d} (x-x_k^S,y-y_k^S)\tilde{d}_{x_i}(x,y)$ is $0$ everywhere). %Note that by symmetry we always have $F(0) = 0$. 
We also see that:

\begin{equation}
\begin{split}
    &\iint_{\mathbb{R}^2}\ n_k \tilde{d} (x-x_k^S,y-y_k^S)\tilde{d}_{y_i}(x,y)\,\mathrm{d}x\,\mathrm{d}y  \\
    & = -n_k \iint_{\mathbb{R}^2}\tilde{d} (\overline{x}-(x_k^S-x_i),\overline{y})  \dot{f}(\sqrt{\overline{x}^2+\overline{y}^2})\frac{\overline{y}}{\sqrt{\overline{x}^2+\overline{y}^2}}\,\mathrm{d}\overline{x}\,\mathrm{d}\overline{y} \\
    & = 0
\end{split}
\end{equation}

since the integrand is an odd function in $\overline{y}$. Returning to the original frame of reference, we see that $\iint_{\mathbb{R}^2}\tilde{d} (x-x_k^S,y-y_k^S)\tilde{d}_{x_i}(x,y)\,\mathrm{d}x\,\mathrm{d}y$ equals $F(\lVert \vec{p}_i - \vec{c}_k \rVert)M_k^{-1}\begin{bmatrix}1 \\ 0\end{bmatrix} = F(\lVert \vec{p}_i - \vec{c}_k \rVert) \overrightarrow{p_i c_k}$, where $\overrightarrow{p_i c_k}$ is a unit vector from $\vec{p}_i$ to $\vec{c}_k$. Consequently,

\begin{equation}
(b) =  -2\sum_{\substack{k = 1}}^{K} n_k F(\lVert \vec{p}_i - \vec{c}_k \rVert) \overrightarrow{p_i c_k}
\end{equation}

Meaning (b) is a sum of vectors from the $i$th agent to each signal center at a distance $V$ or less from it. By the same method, we can compute (c) and conclude that:  

\begin{equation}
    \vec{v}_i(\vec{\textbf{q}}) = 2\Bigg(\sum_{\substack{j = 1}}^{\mathcal{N}} F(\lVert \vec{p}_i - \vec{p}_j \rVert)\overrightarrow{p_i p_j} -\sum_{\substack{k = 1}}^{K} n_k F(\lVert \vec{p}_i - \vec{c}_k \rVert) \overrightarrow{p_i c_k}\Bigg)
    \label{eq:freddyattractionrepulsiondynamics}
\end{equation}

Since at time $t+1$ the $i$th agent updates its position to $\vec{p}_i(t)- \delta  \frac{\vec{v}_i(\vec{\textbf{q}})}{\lVert\vec{v}_i(\vec{\textbf{q}})\lVert}$, we can interpret our agents' dynamics as \textit{attraction-repulsion dynamics}, where the agents are pulled toward the signal centers and pushed away by nearby agents at a magnitude determined by $F(\cdot)$. Since $F(r)$ equals $0$ whenever $r > 2V$, we see that the $i$th agent only needs visibility range $V_A = 2V$ to compute $\vec{v}_i(\vec{\textbf{q}})$. Hence, setting $V = V_A/2$ guarantees that our agents can move according to the dynamics outlined in (\ref{eq:freddydynamics}).

%Given that the dynamics (\ref{eq:freddyattractionrepulsiondynamics}) are determined entirely by $F(\cdot)$, it is convenient, for the sake of designing an effective algorithm, to forget about $f(\cdot)$ and  experiment directly with choices of $F(\cdot)$. The caveat to this approach is that given some arbitrary choice of $F$, a function $f$ that results in $F$ when plugged into Equation (\ref{eq:freddycomputeH}) is not guaranteed to exist. In fact, in the following sections we will give examples of $F(\cdot)$ that explicitly cannot be derived from any choice of $f$ (but can be closely approximated by an appropriate $f$). Hence, this section can be considered a motivating special case of the more general attraction-repulsion dynamics proposed in (\ref{eq:freddyattractionrepulsiondynamics}).

\textbf{Example computation.} Let us provide an explicit computation of $F(\cdot)$. Suppose $V_A = V = \infty$ and $f(r) = e^{-\Lambda r^2}$. Plugging $f$ into (\ref{eq:freddycomputeH}) we see that:

\begin{equation}
\begin{split}
     F(r) 
     & = \iint_{\mathbb{R}^2}{e^{-\Lambda ((\overline{x}-r)^2+\overline{y}^2)} (2\Lambda e^{-\Lambda (\overline{x}^2+\overline{y}^2)}\overline{x})}\mathrm{d}\overline{x} \mathrm{d}\overline{y} \\
     & = 2\Lambda \int_{-\infty}^{\infty}{e^{-2\Lambda\overline{y}^2} \mathrm{d}\overline{y}}\int_{-\infty}^{\infty}{e^{-\Lambda((\overline{x}-r)^2 + \overline{x}^2)}}\overline{x} \mathrm{d} \overline{x} \\
     & =  \sqrt{2\pi\Lambda} \cdot \frac{r\sqrt{\frac{\pi}{2}}e^{-\frac{1}{2}\Lambda r^2}}{2\sqrt{\Lambda}} = \frac{1}{2}\pi r e^{-\frac{1}{2}\Lambda r^2}
\end{split}
\end{equation}

To extend this solution to agents with limited visibility, we can introduce the cut-off point $V_A/2$ to $f$ such that $f(r) = e^{-\Lambda r^2}$ for all $r \leq V_A/2$ and $f(r) = 0$ for all $r > V_A/2$, and recompute (\ref{eq:freddycomputeH}). As discussed above, the resulting function will equal $0$ for all $r > V_A$. When $\Lambda$ is sufficiently large, $F(V_A)$ is very close to $0$, so as a crude approximation we may set the cut-off point $V_A$ directly as part of $F$, such that $F(r) = \frac{1}{2}\pi r e^{-\frac{1}{2}\Lambda r^2}$ for all $r \leq V_A$ and $F(r) = 0$ otherwise. 

\begin{figure}[ht]
    \begin{adjustwidth}{\figureWidthAdjustment}{\figureWidthAdjustment}
     \centering
     \begin{subfigure}[t]{0.32\columnwidth}\hfil
         \centering
         \includegraphics[width=\linewidth]{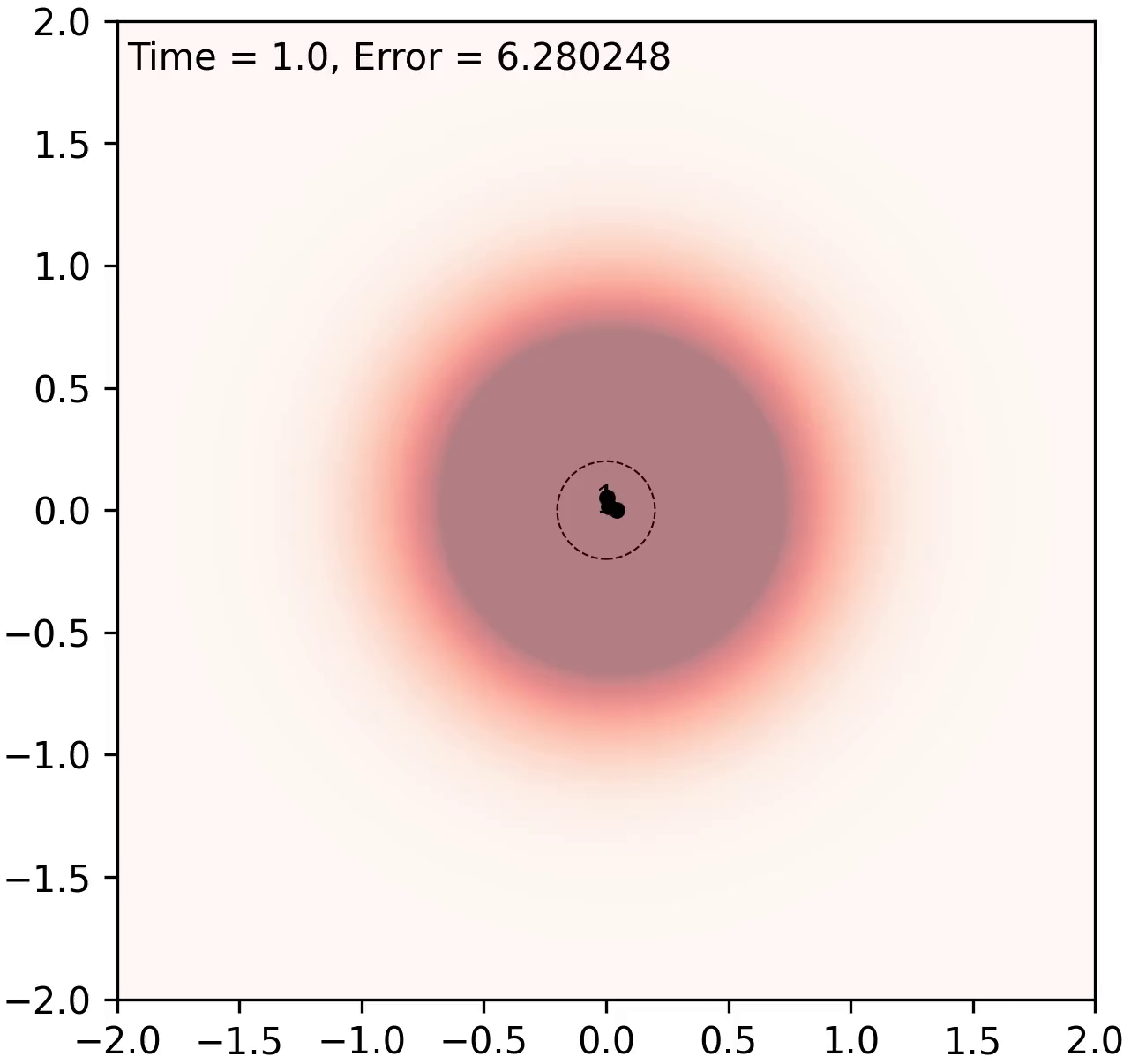}
     \end{subfigure}
     \begin{subfigure}[t]{0.32\columnwidth}\hfil
         \centering
         \includegraphics[width=\linewidth]{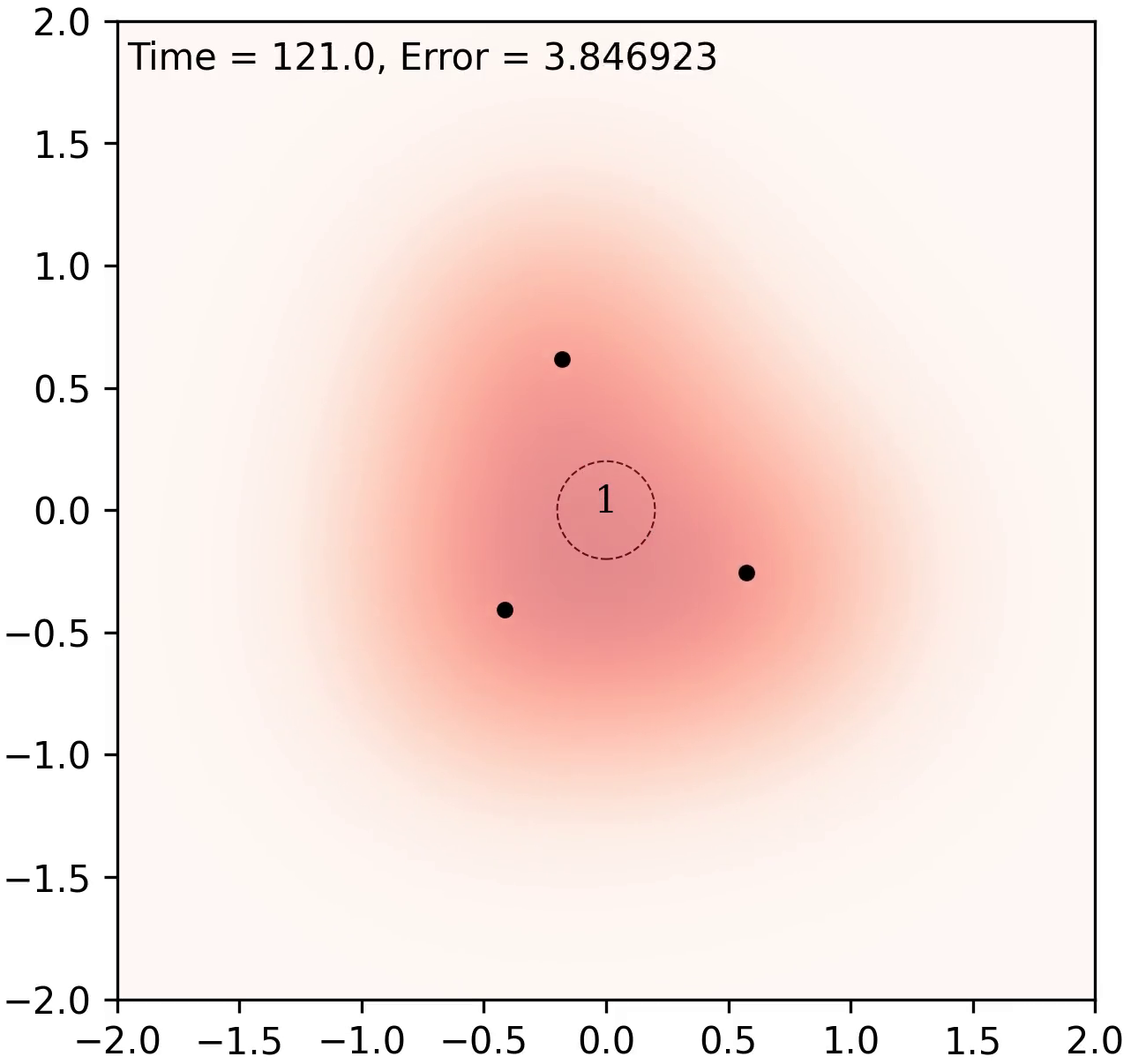}
     \end{subfigure}
     \begin{subfigure}[t]{0.32\columnwidth}\hfil
         \centering
         \includegraphics[width=\linewidth]{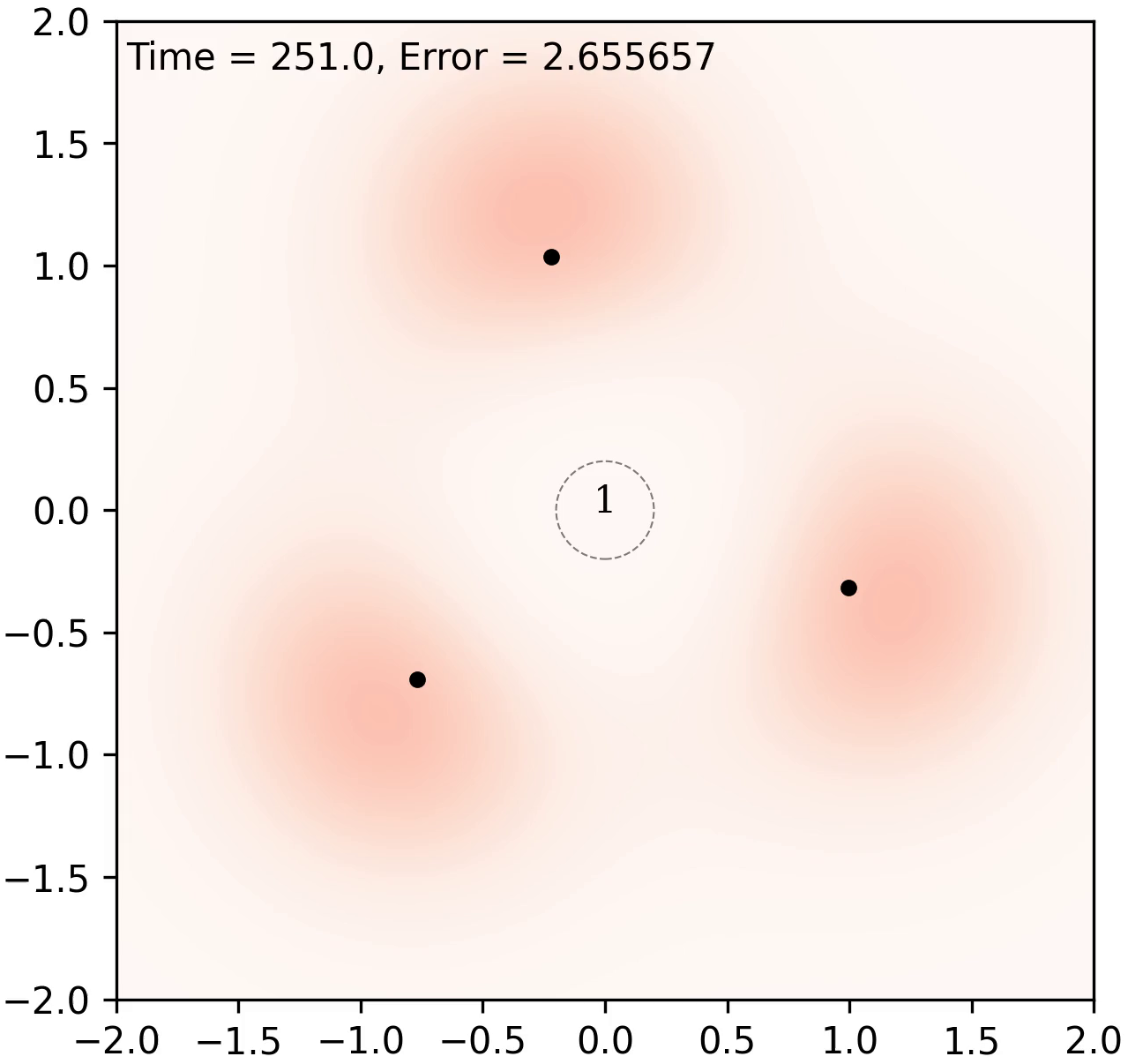}
     \end{subfigure}
     
     \begin{subfigure}[t]{0.32\columnwidth}\hfil
         \centering
         \includegraphics[width=\linewidth]{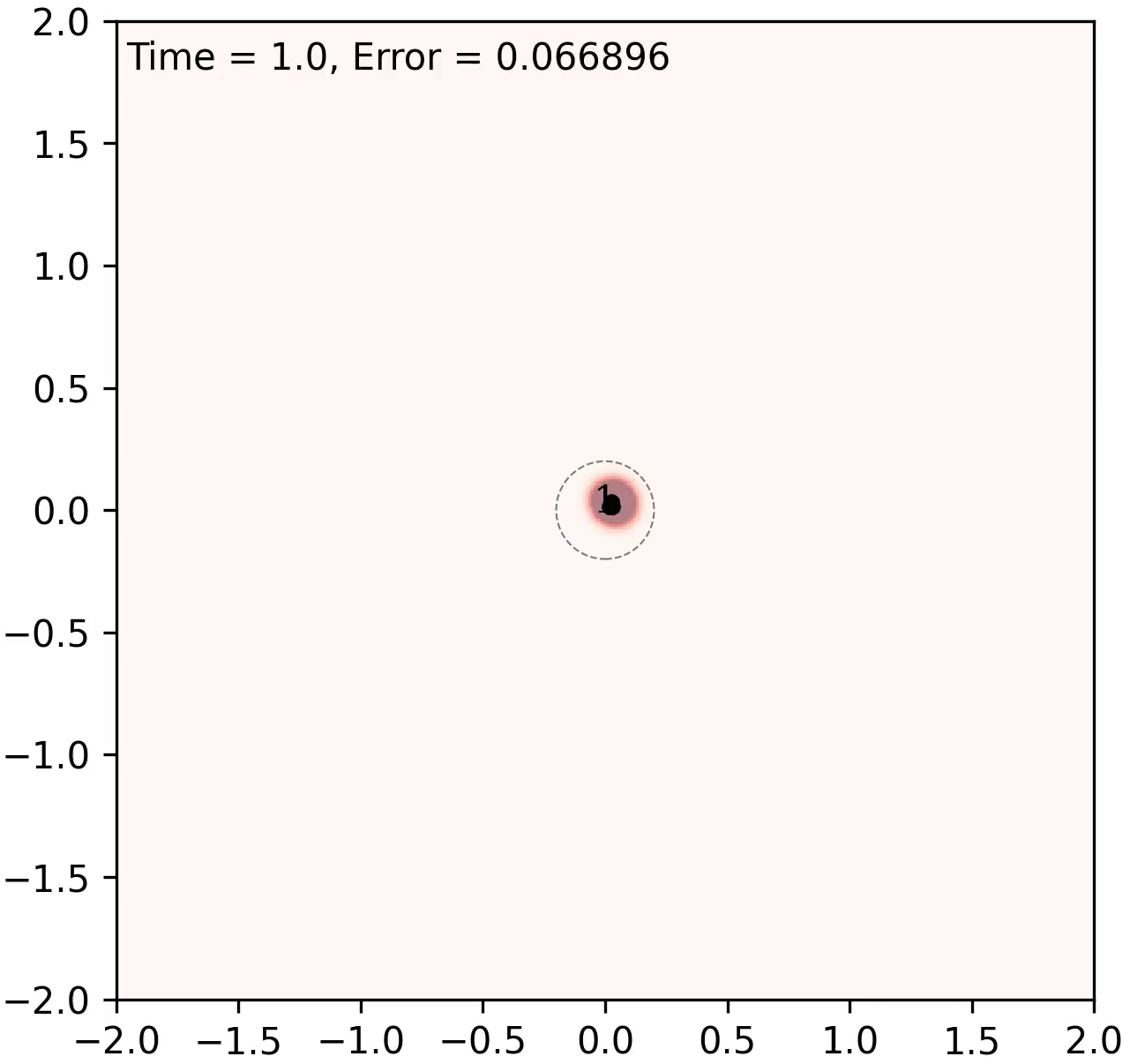}
     \end{subfigure}
     \begin{subfigure}[t]{0.32\columnwidth}\hfil
         \centering
         \includegraphics[width=\linewidth]{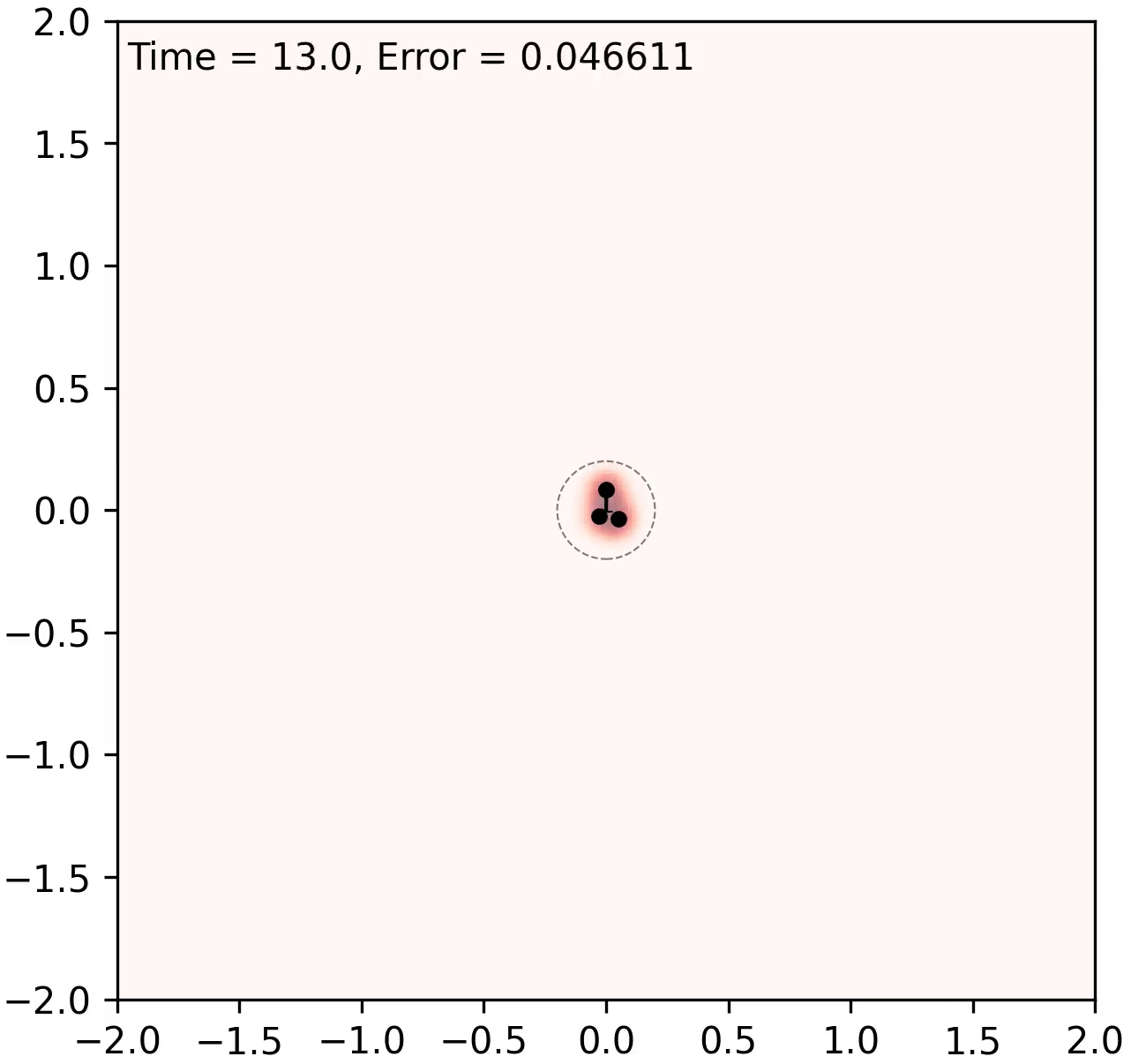}
     \end{subfigure}
     \begin{subfigure}[t]{0.32\columnwidth}
         \centering
         \includegraphics[width=\linewidth]{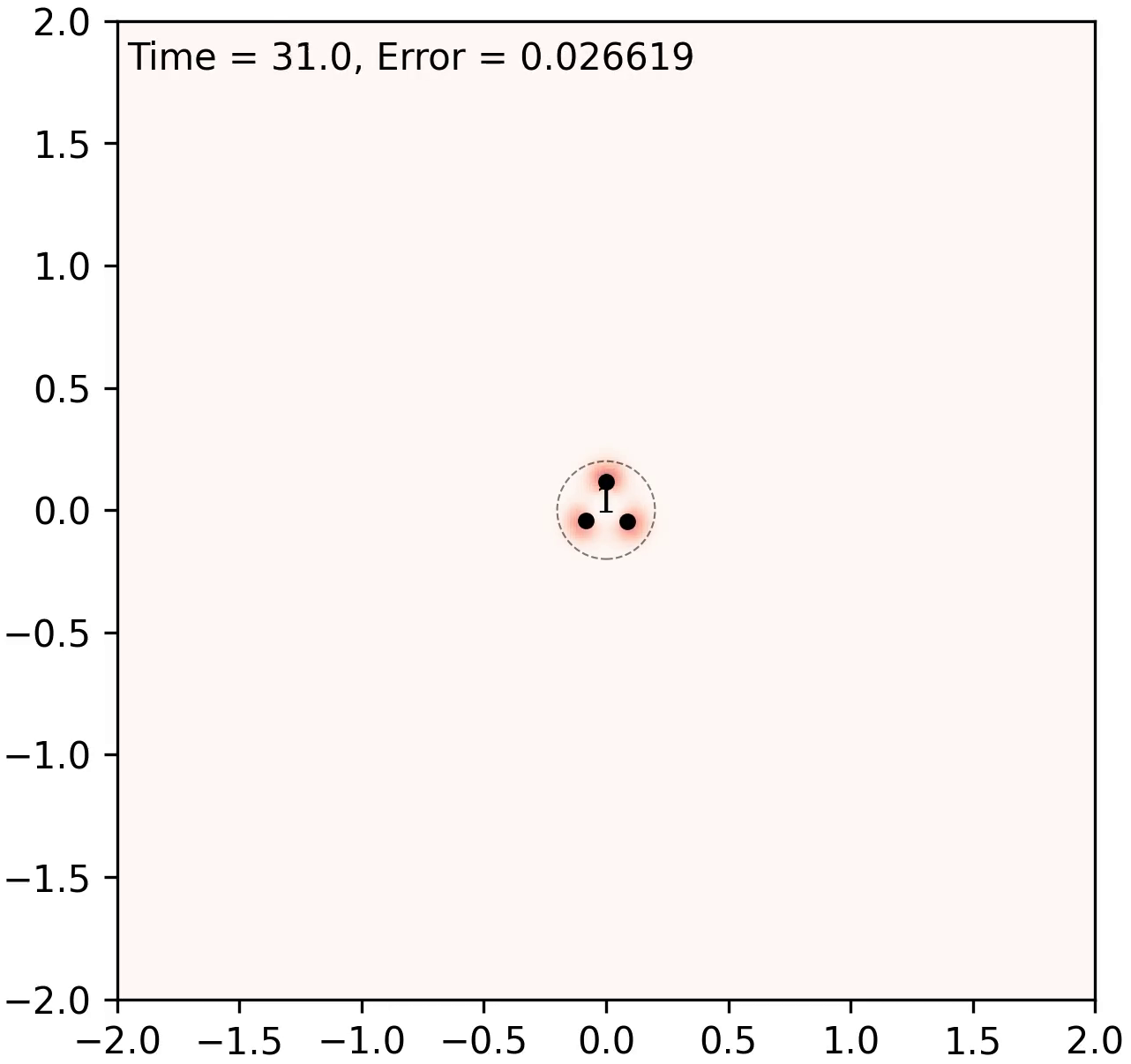}
     \end{subfigure}
     \end{adjustwidth}
     \caption{The top row and bottom row depict two simulations of signal coverage attraction-repulsion dynamics with signal function $f(r) = e^{-\Lambda r^2}$. In the top row $\Lambda = 1$, and in the bottom row $\Lambda = 1000$. In both simulations there is a single  signal center with demand $n_1 = 1$. The heat map illustrates the values of the error function $\Psi$ given the current agent positions. The current time step and value of $\mathcal{G}$ are shown in the top left corner of each frame. The rightmost (i.e., third) frame of each simulation shows the agent formation that minimizes the total error $\mathcal{G}$. Larger values of $\Lambda$ cause the agents to concentrate closer to the signal center. Note that $\Psi$ and $\mathcal{G}$ are different functions in the first and second simulation, since they both depend on $f$.}
     \label{fig:Sim45_freddy_diffusion_example}
\end{figure}

\begin{figure}[ht]
    \begin{adjustwidth}{\figureWidthAdjustment}{\figureWidthAdjustment}
     \centering
     \begin{subfigure}[t]{0.32\columnwidth}\hfil
         \centering
         \includegraphics[width=\linewidth]{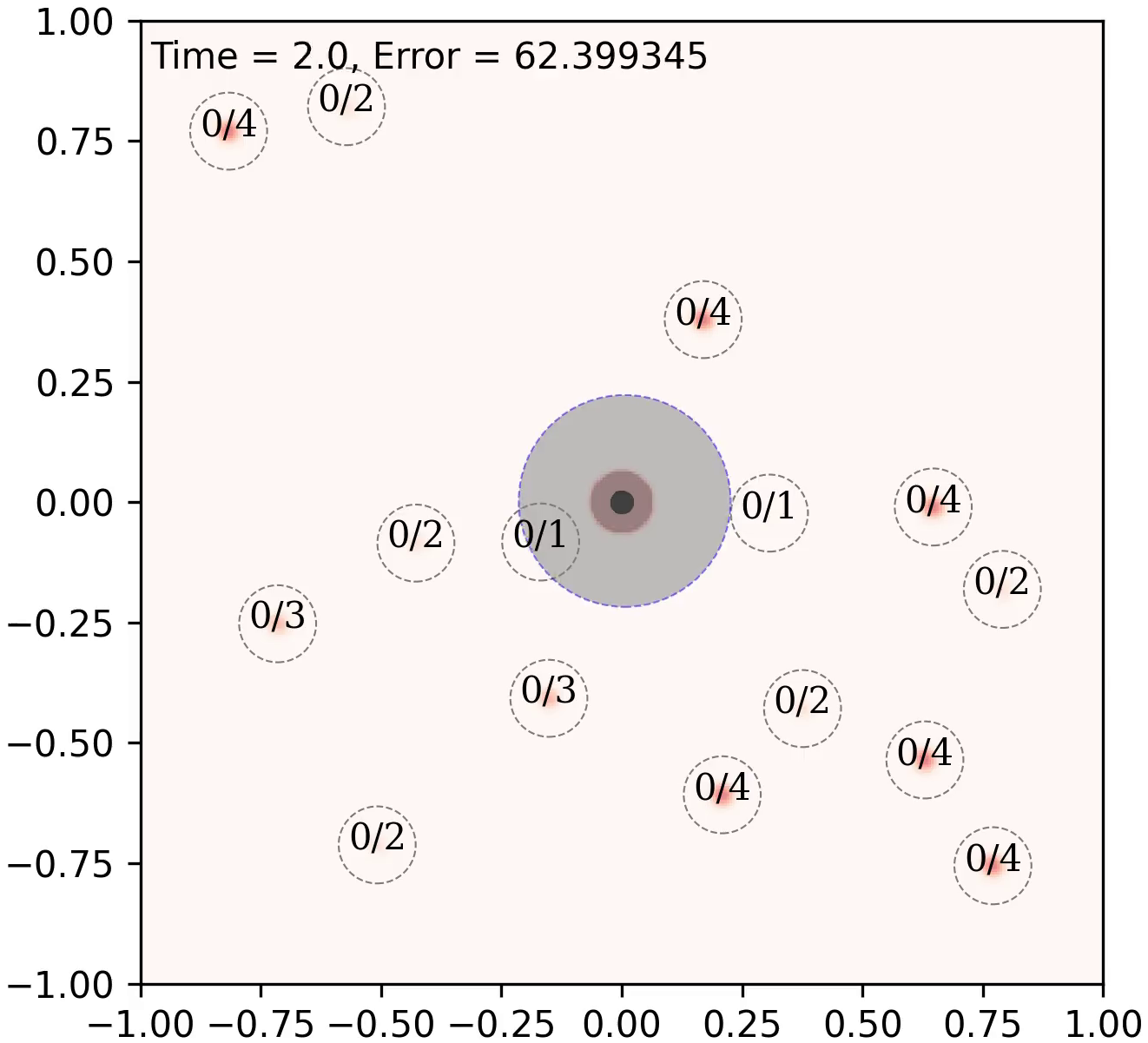}
     \end{subfigure}
     \begin{subfigure}[t]{0.32\columnwidth}\hfil
         \centering
         \includegraphics[width=\linewidth]{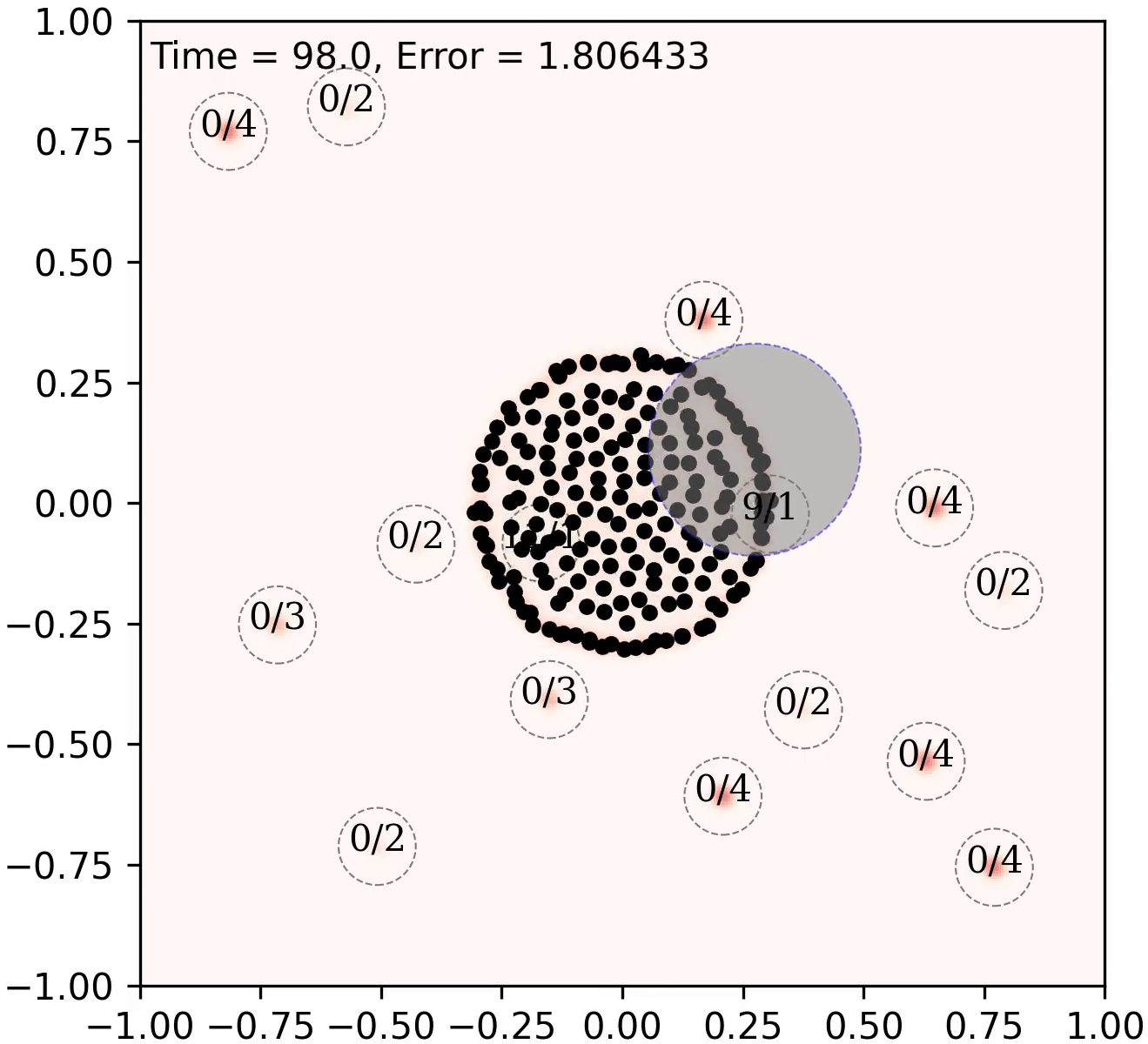}
     \end{subfigure}
     \begin{subfigure}[t]{0.32\columnwidth}\hfil
         \centering
         \includegraphics[width=\linewidth]{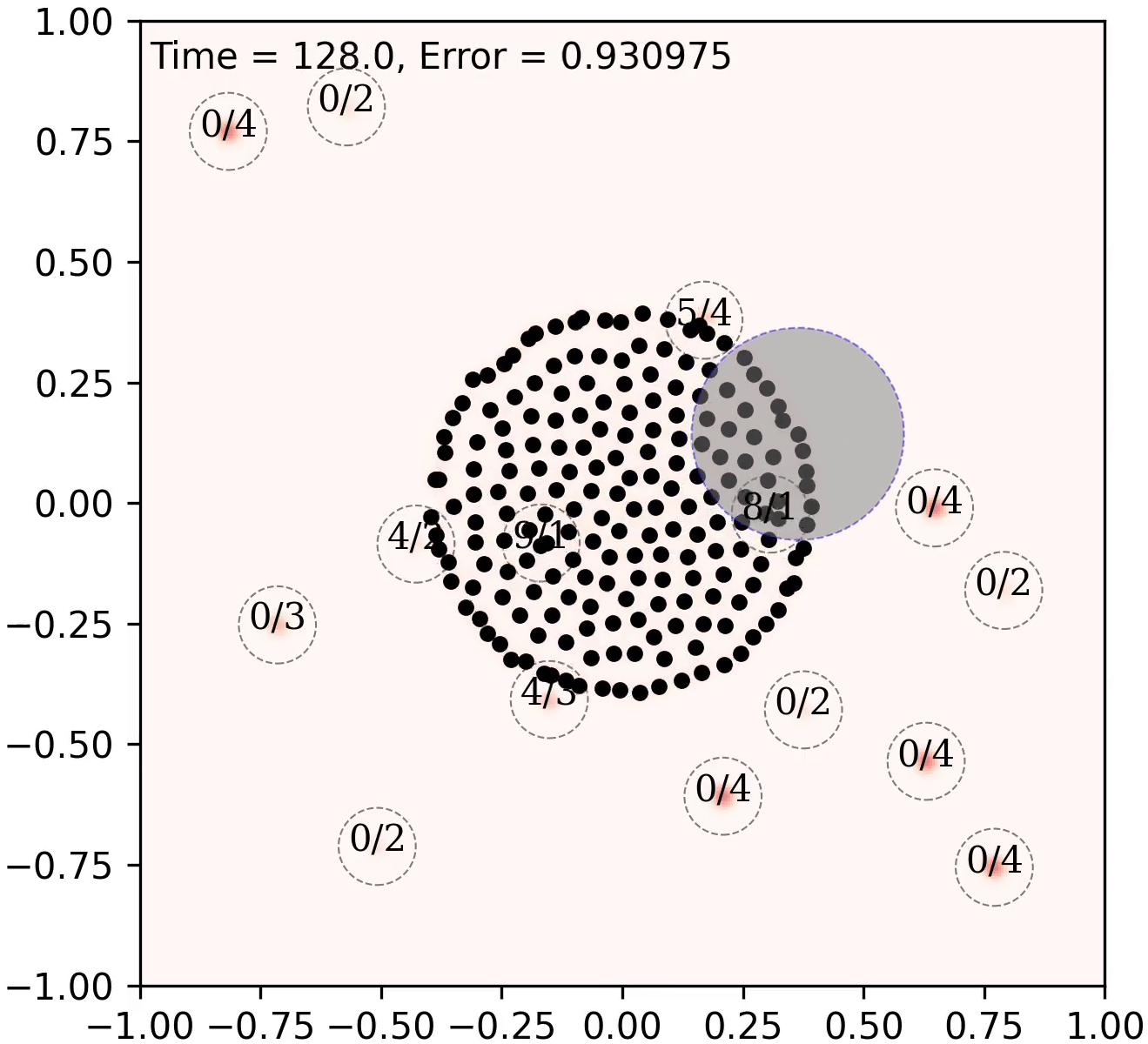}
     \end{subfigure}
     
     \begin{subfigure}[t]{0.32\columnwidth}\hfil
         \centering
         \includegraphics[width=\linewidth]{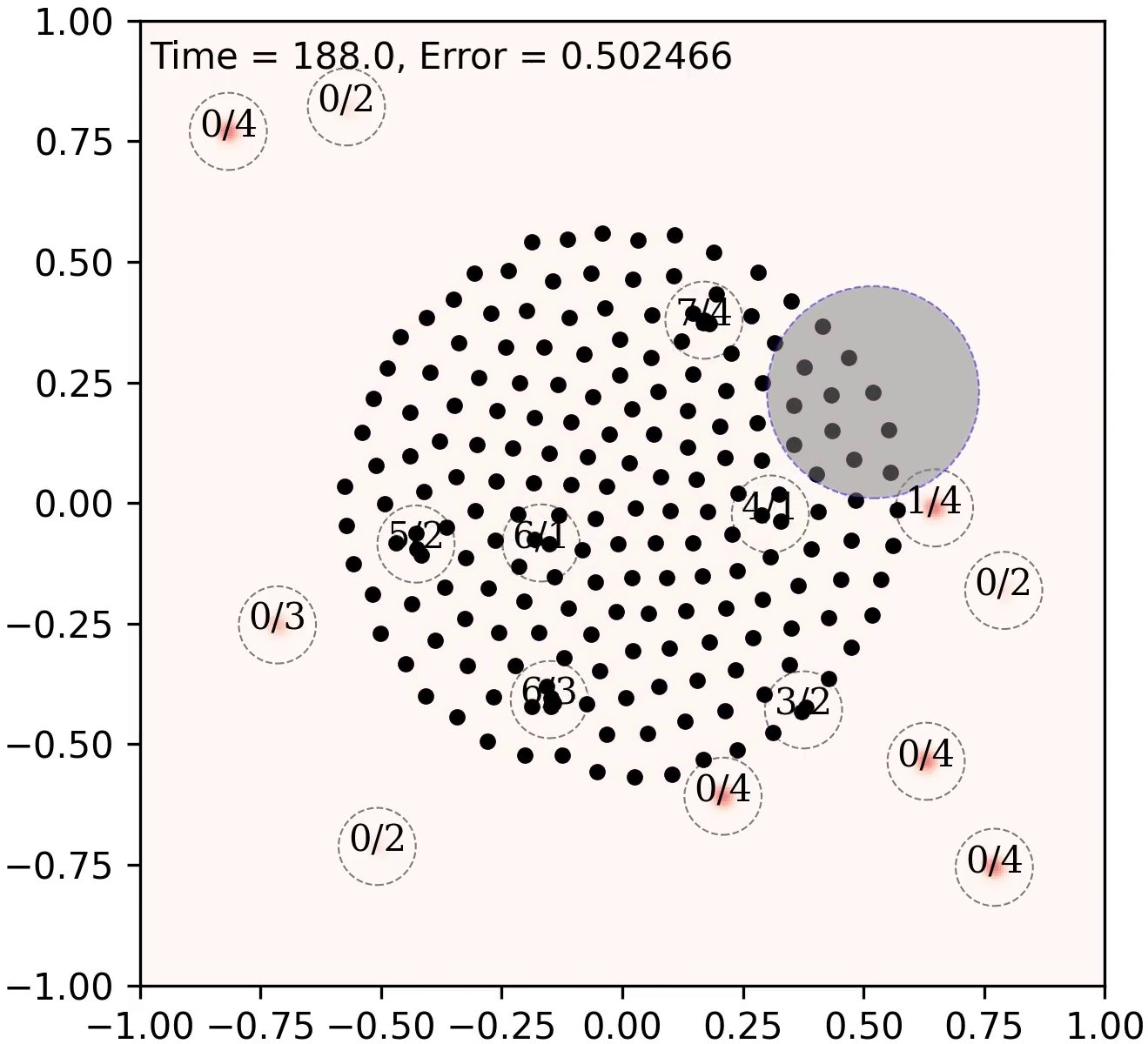}
     \end{subfigure}
     \begin{subfigure}[t]{0.32\columnwidth}\hfil
         \centering
         \includegraphics[width=\linewidth]{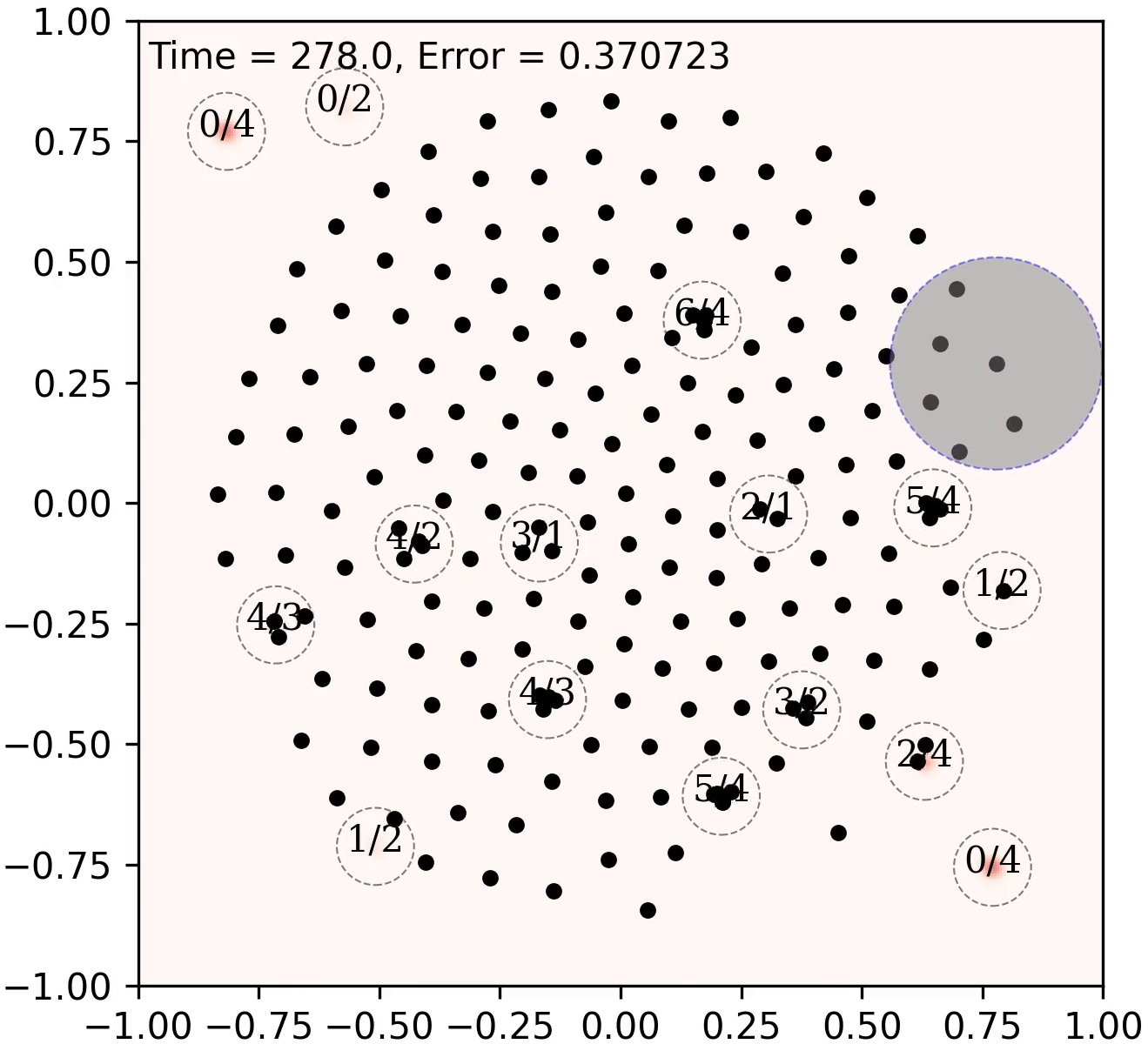}
     \end{subfigure}
     \begin{subfigure}[t]{0.32\columnwidth}
         \centering
         \includegraphics[width=\linewidth]{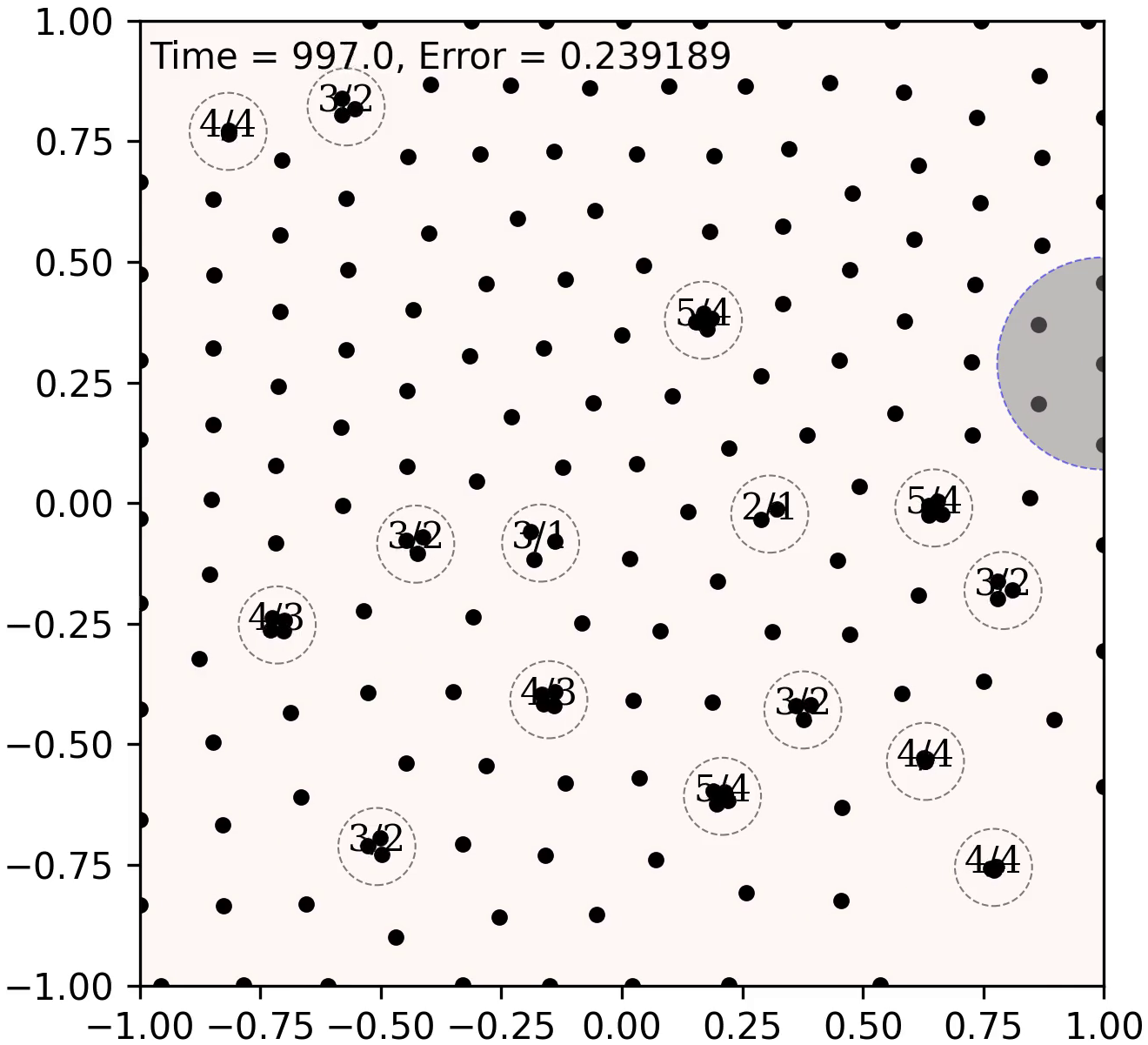}
     \end{subfigure}
     \end{adjustwidth}
     \caption{A simulation of signal coverage attraction-repulsion dynamics for agents with limited sensing range. We set $f(r) = e^{-1000 r^2}$ when $r < V_A/2$ and $f(r) = 0$ otherwise. The numbers on each signal denote the number of agents in its proximity, and the value of $n_i$. The gray disk depicts agents' visibility range. Since $f(r)$ is highly concentrated at $0$, we see that $n_i$ or more agents concentrate at the $i$th signal center, in effect completing a target assignment task. The heat map illustrates the values of the error function $\Psi$ given the current agent positions. The current time step and value of $\mathcal{G}$ are shown in the top left corner of each frame.}
     \label{fig:Sim6_signal_coverage_target_assignment}
\end{figure}

Increasing the parameter $\Lambda$ results in a more concentrated signal function $f$, hence results in  agents concentrating closer to the signal centers $\vec{c}_i$ - see Figure \ref{fig:Sim45_freddy_diffusion_example}. As mentioned earlier in this section,  when $f$ is highly concentrated near $0$, signal coverage can be used for target assignment. This is illustrated in Figure \ref{fig:Sim6_signal_coverage_target_assignment}.

We emphasize that this is just an example, and there are infinite possible choices of $f$ leading to different dynamics. We also note that even when an analytic expression for (\ref{eq:freddycomputeH}) is not available, the agents' dynamics can be efficiently computed by caching numerical approximations of $F(r)$ (since $F(r)$ is a one-dimensional function that equals $0$ outside the interval $[0,V_A]$, it is non-expensive to approximate and cache).

%We emphasize that this is merely a computational example. In simulations, this particular $F(\cdot)$ turns out not to be very effective. In the following sections we give several choices of $F(\cdot)$ that work well in practice.

%If $V_A$ is finite, a possible choice of $F$ is $F(z) =  \frac{1}{2}\pi z e^{-\frac{1}{2}\Lambda z^2}$ for all $z \leq V_A/2$ and $F(z) = 0$ otherwise. 

%We emphasize that these are just examples, and there are infinite possible choices of $f$ and $F$, each leading to different dynamics. For example, we may define $f(r) = 1-r$ for $r \geq 0.5$ and $f(r) = e^{-r^4}$ for $r \leq 0.5$. This results in a discrete change to the agents' dynamics when they get close enough to the center of a target, pulling them in more strongly and making it harder for them to get away, and likewise a strong, discrete repulsion effect when agents get too close to each other.

% and later apply the inverse rotation $\textbf{\mathrm{M}}$ to return to the original frame of reference.

\textbf{Adding random noise.} Depending on $V_A$ and the size of the subregion $\mathcal{L}$, agents may become or initialize isolated--unable to see any other agent or target. An isolated acting on the dynamics outlined in  (\ref{eq:freddydynamics}) doesn't move. This is undesirable: we would like such agents to explore their environment and search for other tasks. To resolve this, we may add a stochastic component to the dynamics such that at every time step, the $i$th agent will move based on the stochastic equation:

\begin{equation}
      \vec{p}_i(t+1) = \vec{p}_i(t) - \delta  \frac{\vec{v}_i(\vec{\textbf{q}}(t))}{\lVert\vec{v}_i(\vec{\textbf{q}}(t))\lVert} + \vec{r}
   \label{eq:freddydynamicsstochastic}
\end{equation}

where $\vec{r}$  is a uniformly random vector of magnitude $\lVert r \rVert \leq \Delta - \delta$. This random step has the effect of breaking deadlocks and causes isolated agents to keep exploring the environment. To prevent agents from wandering too far from $\mathcal{L}$ (which is the region of interest that contains the signals), if (\ref{eq:freddydynamicsstochastic}) would move an agent to a point where $\Phi(x,y) = 0$, the agent instead stays put.

\subsubsection{Signal coverage for nonidentical signals}
\label{section:freddygeneralchoicesofphi}

Instead of having each signal center emit the same (weighted) signal $\tilde{d}$, we may define $\Phi(x,y)$ as a sum of different signals: 

\begin{equation}        
\label{eq:generalized total demand} 
    \Phi(x,y)= 1 +  \sum_{k = 1}^{\mathcal{K}} \tilde{d}_k(x-x_k^S,y-y_k^S) = 1 +  \sum_{k = 1}^{\mathcal{K}} f_k (\lVert \begin{bmatrix}
 x   \\
y  \\
\end{bmatrix} - \vec{c}_k \rVert)
\end{equation}

where $\tilde{d}_k(x,y) = f_k(\lVert \begin{bmatrix}
 x   \\
y  \\
\end{bmatrix} \rVert)$ is a symmetric function representing the environmental signal centred at $\vec{c}_k$, and similar to before we assume $f_k(r) = 0$ for all $r \geq V$. We thus need to compute the gradient (\ref{eq:freddyintegralhalfwaystep}), which can be split into 

\begin{enumerate}[label=(\alph*)]
\item $2\iint_{\mathbb{R}^2}\D\,\mathrm{d}x\,\mathrm{d}y$
\item $ 2\sum_{\substack{k = 1}}^{K}\iint_{\mathbb{R}^2}\tilde{d}_k (x-x_k^S,y-y_k^S)\D\,\mathrm{d}x\,\mathrm{d}y$
\item $-2\iint_{\mathbb{R}^2}\big(\sum_{j=1}^{\mathcal{N}} 
    \tilde{d}(x-x_j,y-y_j)\big)\D\,\mathrm{d}x\,\mathrm{d}y \\$
\end{enumerate}

Carrying out the same analysis as for the non-general case, we see that $(a) = 0$ and $(c) = 2\sum_{\substack{j = 1}}^{\mathcal{N}} F(\lVert \vec{p}_i - \vec{p}_j \rVert)\overrightarrow{p_i p_j}$. Replacing $\tilde{d}$ with $\tilde{d}_k$ in Equation  (\ref{eq:freddycomputeH}), we see that (\ref{eq:freddycomputeH}) equals $-F_k(\lVert \vec{p}_i - \vec{c}_k \rVert)$ for some function $F_k$ which depends on $f_k$, hence $(b) = -2\sum_{\substack{k = 1}}^{K} F_k(\lVert \vec{p}_i - \vec{c}_k \rVert) \overrightarrow{p_i c_k}$. Consequently, gradient descent on $\mathcal{G}$ leads to generalized attraction-repulsion dynamics of the form

\begin{equation}
\begin{split}
    \vec{p}_i(t+1) & = \vec{p}_i(t) - \delta  \frac{\vec{v}_i(\vec{\textbf{q}}(t))}{\lVert\vec{v}_i(\vec{\textbf{q}}(t))\lVert} \\
    \vec{v}_i(\vec{\textbf{q}}) & = \sum_{\substack{j = 1}}^{\mathcal{N}} F(\lVert \vec{p}_i - \vec{p}_j \rVert)\overrightarrow{p_i p_j} -\sum_{\substack{k = 1}}^{K} F_k(\lVert \vec{p}_i - \vec{c}_k \rVert) \overrightarrow{p_i c_k} 
\end{split}
\label{eq:attractionrepulsiondynamicsgoalgeneral}
\end{equation}

When $f_k = n_k f$ we get that $F_k = n_kF$, making (\ref{eq:attractionrepulsiondynamicsgoalgeneral}) equal  (\ref{eq:attractionrepulsiondynamicsgoal}).

%Let us note that the functions $F(\cdot)$ and $F_k(\cdot)$ are not arbitrarily chosen functions - they explicitly depend on the signals $f_k$ and $f$, and are computed via an integral of the form (\ref{eq:freddycomputeH}). It is interesting and useful, however, to consider attraction-repulsion dynamics in more generality, divorced from such constraints. We explore this topic in  Section \ref{section:generalizedattractionrepulsion}.

\subsection{Generalized attraction-repulsion dynamics}
\label{section:generalizedattractionrepulsion}

In the previous subsection we gave a method for multi-agent signal coverage based on attraction-repulsion dynamics of the form (\ref{eq:attractionrepulsiondynamicsgoalgeneral}). We showed that these dynamics arise naturally from gradient descent over an appropriate $\Psi$ function. As an alternative way of thinking about these dynamics, we could relate them to a general form of attraction-repulsion dynamics similar to that which appears in \cite{gazipasino2004stability}, where at every time step the $i$th agent moves according to the rule of motion:

\begin{equation}
\begin{split}
    \vec{p}_i(t+1) & = \vec{p}_i(t) - \delta  \frac{\vec{v}_i(\vec{\textbf{q}}(t))}{\lVert\vec{v}_i(\vec{\textbf{q}}(t))\lVert} \\
    \vec{v}_i(\vec{\textbf{q}}) & = \sum_{\substack{j = 1}}^{\mathcal{N}} h(\lVert \vec{p}_i - \vec{p}_j \rVert)\overrightarrow{p_i p_j} - \frac{\partial }{\partial \vec{p}_i}\Phi(\vec{p_i})
\end{split}
    \label{eq:attractionrepulsiondynamicsgoalgeneralsection}
\end{equation}

in which $h(r) : \mathbb{R}_{\geq 0} \to \mathbb{R}$ is some function and $0 < \delta  \leq \Delta$ is a constant. We can understand $h$ as a repulsive force between agents, and the gradient of the demand profile $\Phi$ as an attractive force: at every time step the $i$th agent attempts to climb the gradient of $\Phi$, but is repulsed by the $j$th agent by an amount dependent on their distance. Because the agents' sensing range is $V_A$, we assume that $h(r) = 0$ for all $r > V_A$. 

The goal of this section is (i) to note that the dynamics (\ref{eq:attractionrepulsiondynamicsgoalgeneral}) are a special case of (\ref{eq:attractionrepulsiondynamicsgoalgeneralsection}) and (ii) to propose several alternative approaches to task allocation that can be recovered from this general form.

%The dynamics of Equation (\ref{eq:attractionrepulsiondynamicsgoalgeneral}) are derived from gradient descent over the signal coverage error function $ \big(\Phi(x,y) - \sum_{\substack{i = 1}}^{\mathcal{N}} \tilde{d} (x-x_i,y-y_i) \big)^2$. Due to the constraints over

%In the other direction, the rule of motion (\ref{eq:attractionrepulsiondynamicsgoalgeneralsection}) cannot necessarily be recovered from (\ref{eq:attractionrepulsiondynamicsgoalgeneral}) for every choice of $h$ and $\Phi$. For example, suppose that $h(r) = 1$ when $r \leq V_A$ and $h(r) = 0$ otherwise; and $\Phi(x,y) = C$ for some constant $C$. Since necessarily $F(0) = F_k(0) = 0$, we see that no choice of $f$ and $f_k$ can make (\ref{eq:attractionrepulsiondynamicsgoalgeneralsection}) equal (\ref{eq:attractionrepulsiondynamicsgoalgeneral}) in this case. It is possible that there exist $f$ and $f_k$ that

The dynamics of Equation (\ref{eq:attractionrepulsiondynamicsgoalgeneral}) are derived from gradient descent over the total error $\iint_{\mathbb{R}^2}\Psi(x,y,\vec{\textbf{q}}) \,\mathrm{d}x\,\mathrm{d}y$ where $\Psi(x,y,\vec{\textbf{q}}) = \big(\Phi(x,y) - \sum_{\substack{i = 1}}^{\mathcal{N}} \tilde{d} (x-x_i,y-y_i) \big)^2$. We would be remiss if we didn't ask whether the dynamics (\ref{eq:attractionrepulsiondynamicsgoalgeneralsection}) similarly minimize some error function, and whether they can similarly be justified through gradient descent. In other words, we would like to know if there exists $\Psi$ such that the gradient $\frac{\partial \mathcal{G}}{\partial \vec{p}_i}(\vec{\textbf{q}})$ of the total error $\mathcal{G}(\vec{\textbf{q}}) = \iint_{\mathbb{R}^2}\Psi(x,y,\vec{\textbf{q}}) \,\mathrm{d}x\,\mathrm{d}y$ at $\vec{\textbf{q}}$ equals the vector  $\vec{v}_i(\vec{\textbf{q}})$ in Equation  (\ref{eq:attractionrepulsiondynamicsgoalgeneralsection}). It turns out such a $\Psi$ does exist: inspired by an idea of \cite{slotineschwager2011unifying}, let us consider the error

\begin{equation}
    \Psi(x,y,\vec{\textbf{q}}) = \big(\frac{1}{2}\sum_{\substack{ j = 1}}^{\mathcal{N}} H (\lVert  \begin{bmatrix}x\\y\end{bmatrix} - \vec{p}_j \rVert) - \Phi(x,y) 
    \big) \sum_{i=1}^{\mathcal{N}} \updelta(x - x_i) \updelta(y - y_i)
    \label{eq:generalattractionrepulsionpsi}
\end{equation}

where $\updelta$ is the Dirac delta function and $H$ is any almost everywhere differentiable function. The total error which the agents seek to minimize under this function $\mathcal{G}(\vec{\textbf{q}})$ is:

\begin{equation}
\begin {split}
    \hspace*{-0.4cm} & \iint_{\mathbb{R}^2}\big(  \frac{1}{2}\sum_{\substack{j = 1}}^{\mathcal{N}} H (\lVert  \begin{bmatrix}x\\y\end{bmatrix} - \vec{p}_j \rVert) -\Phi(x,y) \big) \sum_{i=1}^{\mathcal{N}} \updelta(x - x_i) \updelta(y - y_i)  \,\mathrm{d}x\,\mathrm{d}y \\
   % \hspace*{-0.6cm} & =\sum_{i=1}^{\mathcal{N}} \int_{-\infty}^{\infty} \int_{-\infty}^{\infty} \big(\frac{1}{2}\sum_{\substack{j = 1}}^{\mathcal{N}} H (\lVert  \begin{bmatrix}x\\y\end{bmatrix} - \vec{p}_j \rVert) - \Phi(x,y) \big) \delta(x - x_i) \delta(y - y_i)
  %  \,\mathrm{d}x\,\mathrm{d}y \\
  %  \hspace*{-0.6cm}& =  \sum_{i=1}^{\mathcal{N}}  \int_{-\infty}^{\infty} \big(  \frac{1}{2}\sum_{\substack{j = 1}}^{\mathcal{N}} H (\lVert  \begin{bmatrix}x_i\\y\end{bmatrix} - \vec{p}_j \rVert) - \sum_{i=1}^{\mathcal{N}}\Phi(x_i,y)  \big)  \delta(y - y_i)
  %   \,\mathrm{d}y \\
     \hspace*{-0.4cm}& =  \frac{1}{2} \sum_{1 \leq i,j \leq \mathcal{N}} H (\lVert \vec{p_i} - \vec{p_j} \rVert) - \sum_{\substack{i = 1}}^{\mathcal{N}}\Phi(\vec{p_i})
\end{split}
\end{equation}

hence the gradient at $\vec{\textbf{q}}$ is 

%\frac{\partial }{\partial \vec{p}_i}\Phi(\vec{p_i}) -  \sum_{1 \leq i \leq j \leq \mathcal{N}} \frac{\partial }{\partial \vec{p}_i}H(\lVert \vec{p}_i - \vec{p}_j \rVert) \\
    
\begin{equation}
\begin{split}
        \frac{\partial \mathcal{G}}{\partial \vec{p}_i}(\vec{\textbf{q}}) & =    \sum_{1 \leq j \leq \mathcal{N}}  \dot{H} (\lVert \vec{p}_i - \vec{p}_j \rVert)  \overrightarrow{p_i p_j} - \frac{\partial }{\partial \vec{p}_i}\Phi(\vec{p_i})
\end{split}
    \label{eq:generalattractionrepulsiongradient}
\end{equation}

Defining $h(r) = \dot{H}(r)$, we see that the dynamics  (\ref{eq:attractionrepulsiondynamicsgoalgeneralsection}) are precisely discrete gradient descent dynamics over $\mathcal{G}$.

We now discuss several task allocation strategies that can be recovered as special cases of (\ref{eq:attractionrepulsiondynamicsgoalgeneralsection}), i.e., through gradient descent over $\iint_{\mathbb{R}^2}\Psi(x,y,\vec{\textbf{q}}) \,\mathrm{d}x\,\mathrm{d}y$ for appropriate choices of $H$ and $\Phi$.

\subsubsection{Signal coverage}

We ask whether the dynamics  (\ref{eq:attractionrepulsiondynamicsgoalgeneral}) we derived for signal coverage can be recovered as a special case. Recall that (\ref{eq:attractionrepulsiondynamicsgoalgeneral})   depends on two integrable functions: $F$ and $F_k$. Let us set $H(r) = \int_{0}^{r} F(x) dx$, $H_k(r) = \int_{0}^{r} F_k(x) \,\mathrm{d}x$ and $\Phi(x,y) = \sum_{\substack{k = 1}}^{K}  H_k(\lVert \begin{bmatrix}x \\ y\end{bmatrix} - \vec{c}_k \rVert)$. We compute that

\begin{equation}
\begin{split}
    \frac{\partial }{\partial \vec{p}_i}\Phi(\vec{p_i}) & = \sum_{\substack{k = 1}}^{K}  \frac{\partial }{\partial \vec{p}_i}H_k(\lVert \vec{p_i} - \vec{c}_k \rVert) \\
    & = \sum_{\substack{k = 1}}^{K}  F_k(\lVert \vec{p_i}- \vec{c}_k \rVert) \overrightarrow{p_i c_k}\textrm{, which implies}
\end{split}
\end{equation}

\begin{equation}
\begin{split}
    (\ref{eq:generalattractionrepulsiongradient})   =  \sum_{\substack{j = 1}}^{\mathcal{N}} F(\lVert \vec{p}_i - \vec{p}_j \rVert)\overrightarrow{p_i p_j} - \sum_{\substack{k = 1}}^{K}  F_k(\lVert \vec{p_i}- \vec{c}_k \rVert) \overrightarrow{p_i c_k}
\end{split}
\end{equation}

hence, for these choices of $H$ and $\Phi$, (\ref{eq:attractionrepulsiondynamicsgoalgeneralsection}) precisely equals  (\ref{eq:attractionrepulsiondynamicsgoalgeneral}).

\subsubsection{Electrostatic target assignment and Coulomb's law}
\label{section:electrostatictargetassignment}
%Recall that in the target assignment problem there are $K$ targets at a priori unknown locations $\vec{c}_1, \ldots \vec{c}_K \in \mathcal{L}$, and for every $1 \leq k \leq K$, we want to bring $n_k$ agents to the location $\vec{c}_k$. The targets' locations are not known to the agents in advance, and targets can only be detected by an agent when they are at a distance $V_A$ or less from it.

In Section \ref{section:signalcoveragemodel} we described the target assignment problem, and noted that target assignment problems can be represented as signal coverage problems where $\Phi$ is highly concentrated  at each target location. This led to us to propose signal coverage as a method for target assignment. Another approach to target assignment is based on Coulomb's law, which states the attraction or repulsion forces between two  particles with electric charges $q_1$ and $q_2$ respectively is $\frac{kq_1q_2}{r^2}$, where $r$ is the distance between the particles, and $k$ is a scaling factor. The idea is to assign each agent a positive charge, creating repulsion between pairs of agents, and each stationary target task a negative charge, creating attraction between agents and targets. Such an approach to particle control is widely used in image halftoning  \cite{gwosdek2014fast,schmaltz2010electrostatic,teuber2011dithering}, in multi-robot collison avoidance and search and rescue \cite{baxter2007multirobotelectrostatic}, and more. Here we discuss this approach in the context of target assignment for large swarms of agents with limited visibility, and show that it is a special case of  (\ref{eq:attractionrepulsiondynamicsgoalgeneral}). 

We begin with the case where $V_A = \infty$, i.e., agents can sense other agents and targets at any distance. Let us define $H(r) = -\frac{1}{r}$ and  $\Phi(x,y) = \sum_{k=1}^K n_k H(\lVert \begin{bmatrix}x\\y\end{bmatrix} - \vec{c}_k \rVert)$.
%\begin{equation*}
%\Phi(x,y) = \sum_{k=1}^K n_k H(\lVert \begin{bmatrix}x\\y\end{bmatrix} - \vec{c}_k \rVert)
%\end{equation*}
Inserting these expressions into (\ref{eq:generalattractionrepulsiongradient}) we get

\begin{equation}
\begin{split}
        \frac{\partial \mathcal{G}}{\partial \vec{p}_i} (\vec{\textbf{q}})& = \sum_{1 \leq i \leq j \leq \mathcal{N}}  \dot{H} (\lVert \vec{p}_i - \vec{p}_j \rVert)  \overrightarrow{p_i p_j} -  \sum_{k=1}^{K} n_k \dot{H} (\lVert \vec{p}_i - \vec{c}_k \rVert)\overrightarrow{p_i c_k}   \\
        & =     \sum_{1  \leq j \leq \mathcal{N}}  \frac{\overrightarrow{p_i p_j}}{\lVert \vec{p}_i - \vec{p}_j \rVert^2} - \sum_{k=1}^{K} n_k \frac{\overrightarrow{p_i c_k}}{\lVert \vec{p}_i - \vec{c}_k \rVert^2}
\end{split}
\label{eq:electrostaticdynamicsinfinitevisibility}
\end{equation}

The dynamics (\ref{eq:electrostaticdynamicsinfinitevisibility}) result from Coulomb's law acting on targets with a positive charge of magnitude $n_k$ and agents with a negative charge of magnitude $1$. These dynamics require infinite sensing range to be computed by the agents; if $V_A$ is finite, we may define a cut-off point for $H$, making it so that agents are not affected by other agents and targets at distance greater than $V_A$: 

\begin{equation*}
H_{V_A}(r) =  
 \begin{cases} 
      -\frac{1}{r} & {r \leq V_A} \\
       0 & {r > V_A}
   \end{cases}
\end{equation*}

This results in dynamics that can be computed by agents with visibility range $V_A$.

Let us briefly relate the electrostatic target assignment approach to our previous, signal coverage-based approach: note that if we could find a signal $f$ for which the integral (\ref{eq:freddycomputeH}) equalled $H_{V_A}(r)$, plugging this $f$ into the attraction-repulsion dynamics we used to solve signal coverage, (\ref{eq:attractionrepulsiondynamicsgoal}), would precisely yield  (\ref{eq:electrostaticdynamicsinfinitevisibility}). However, it seems difficult to find such an $f$. Hence, although (\ref{eq:electrostaticdynamicsinfinitevisibility}) is of a very similar form to (\ref{eq:attractionrepulsiondynamicsgoal}), we leave open the question of whether it can be recovered as a special case of signal coverage.

\begin{figure}[!ht]
    \begin{adjustwidth}{\figureWidthAdjustment}{\figureWidthAdjustment}
     \centering
     \begin{subfigure}[t]{0.32\columnwidth}\hfil
         \centering
         \includegraphics[width=\linewidth]{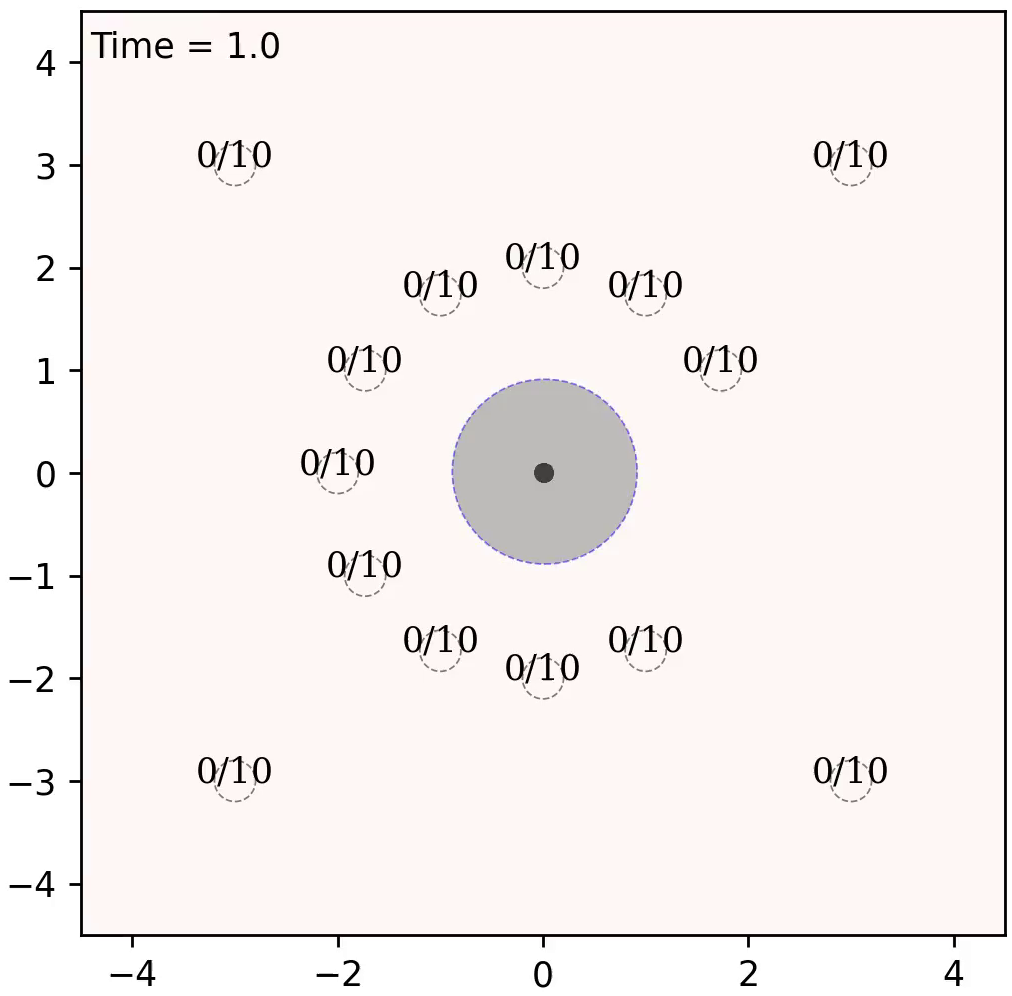}
     \end{subfigure}
     \begin{subfigure}[t]{0.32\columnwidth}\hfil
         \centering
         \includegraphics[width=\linewidth]{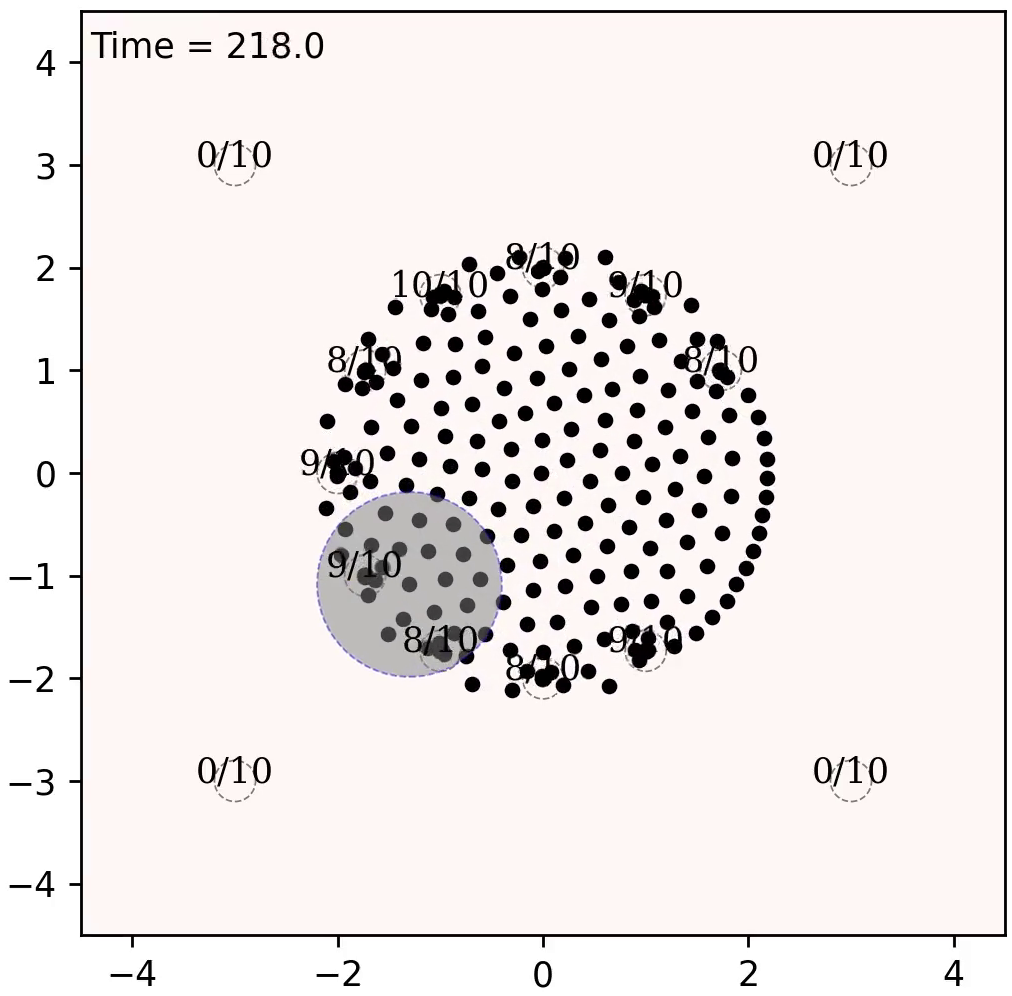}
     \end{subfigure}
     \begin{subfigure}[t]{0.32\columnwidth}\hfil
         \centering
         \includegraphics[width=\linewidth]{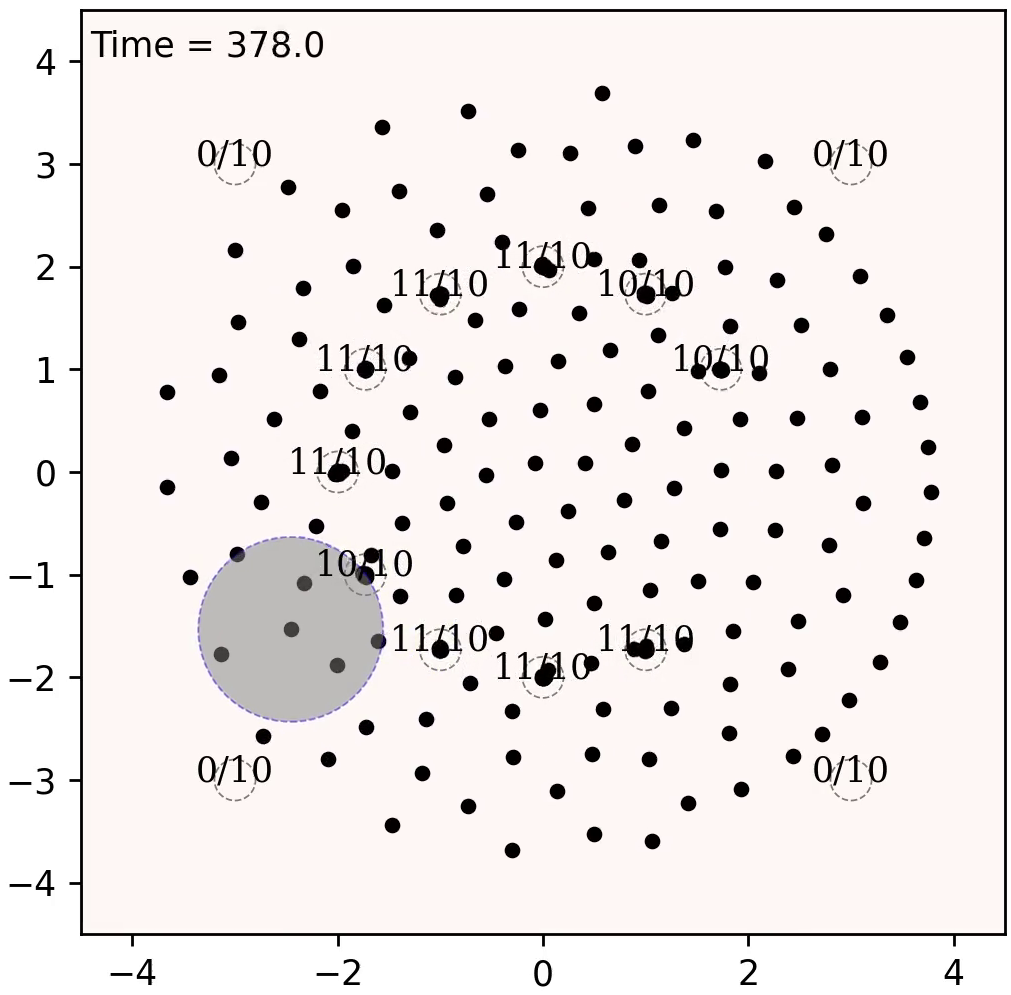}
     \end{subfigure}
     
     \begin{subfigure}[t]{0.32\columnwidth}\hfil
         \centering
         \includegraphics[width=\linewidth]{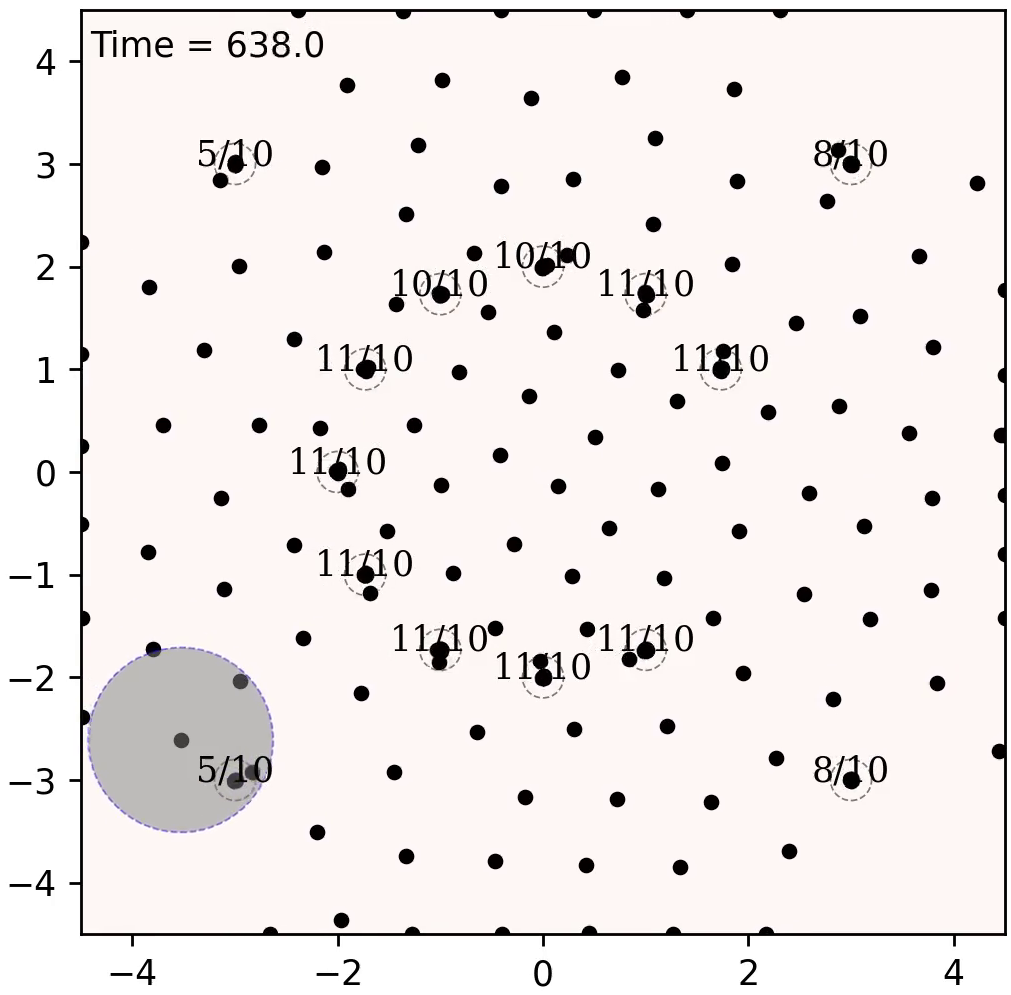}
     \end{subfigure}
     \begin{subfigure}[t]{0.32\columnwidth}\hfil
         \centering
         \includegraphics[width=\linewidth]{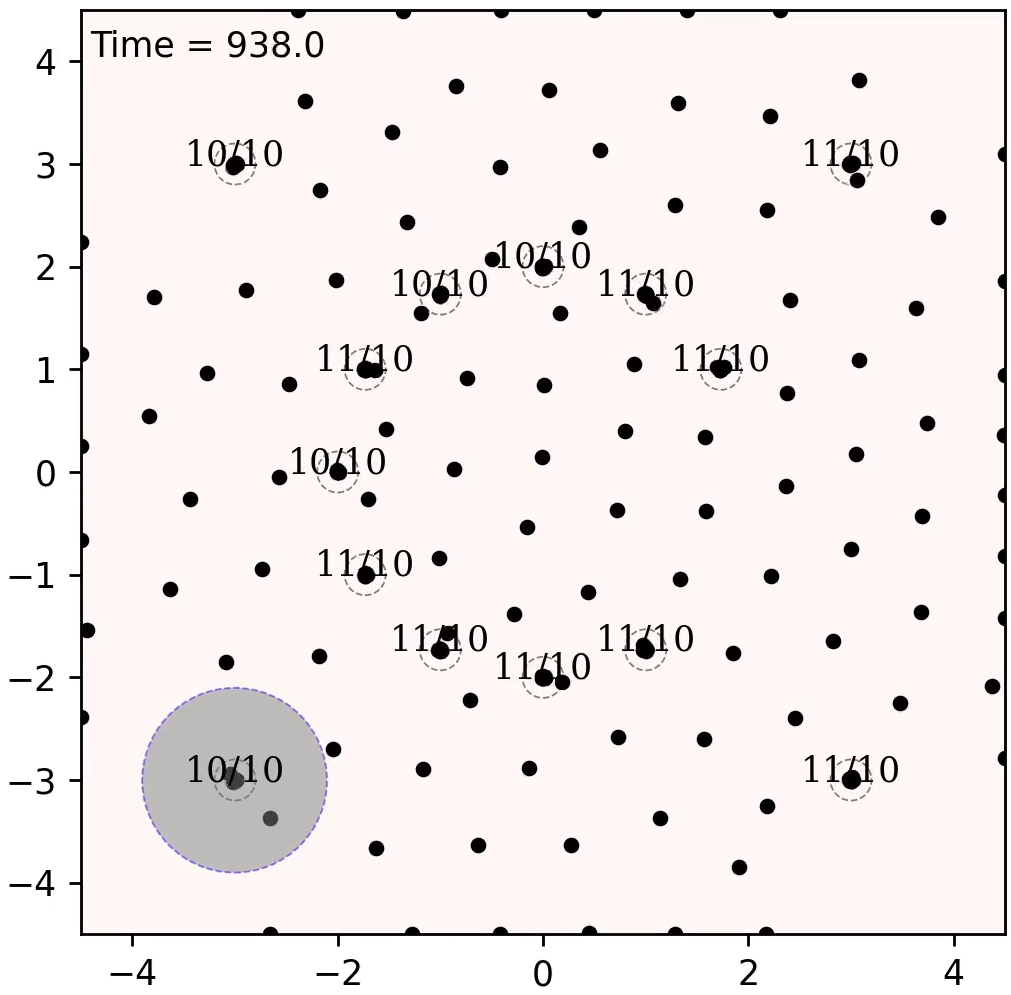}
     \end{subfigure}
     \begin{subfigure}[t]{0.32\columnwidth}
         \centering
         \includegraphics[width=\linewidth]{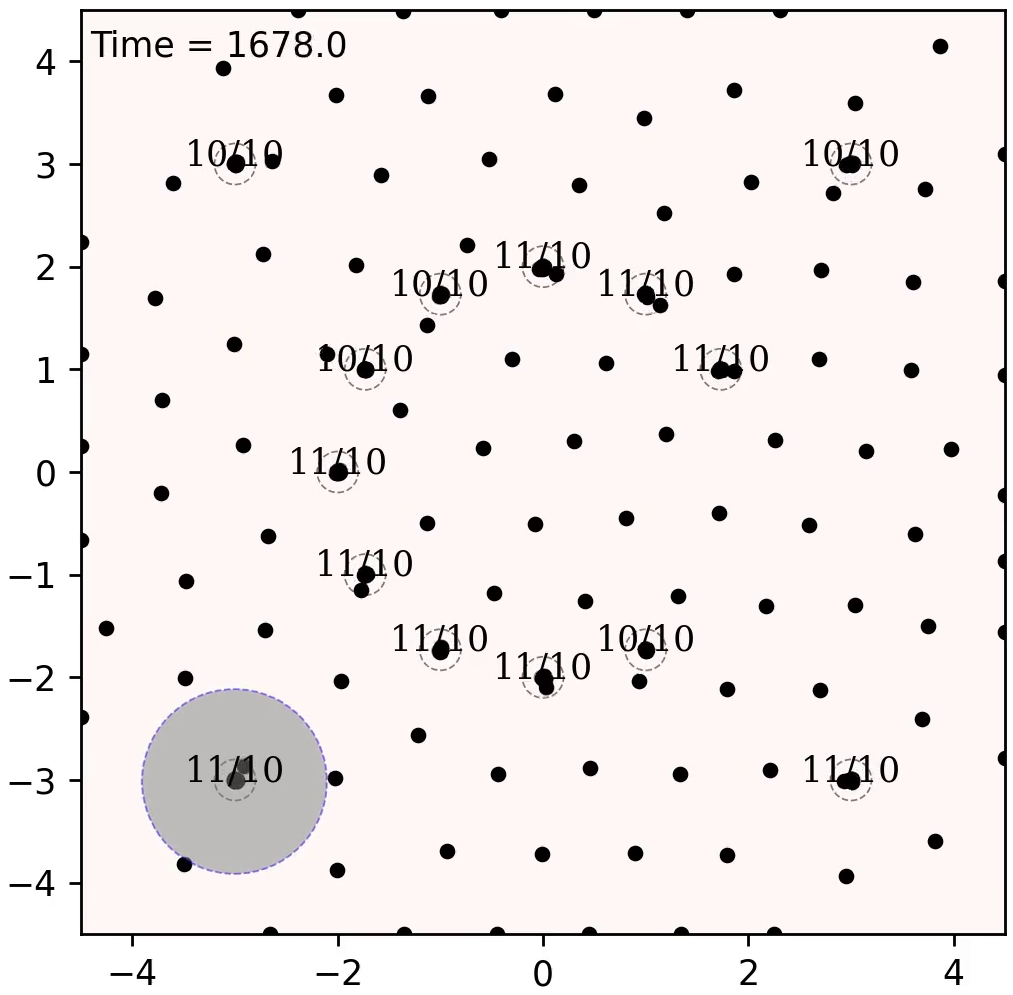}
     \end{subfigure}
     \end{adjustwidth}
     \caption{Limited sensing range electrostatic target assignment. The number of agents demanded by each target and the  number of agents in each target's proximity are depicted. Agent sensing range is illustrated by the gray disk. There are 250 agents, and the total demand of the targets is 140. Agents are all initiated at the center of the region.}
     \label{fig:Sim2_electrostatic_target_assignment}
\end{figure}

\begin{figure}[ht]
    \begin{adjustwidth}{\figureWidthAdjustment}{\figureWidthAdjustment}
     \centering
     \begin{subfigure}[t]{0.32\columnwidth}\hfil
         \centering
         \includegraphics[width=\linewidth]{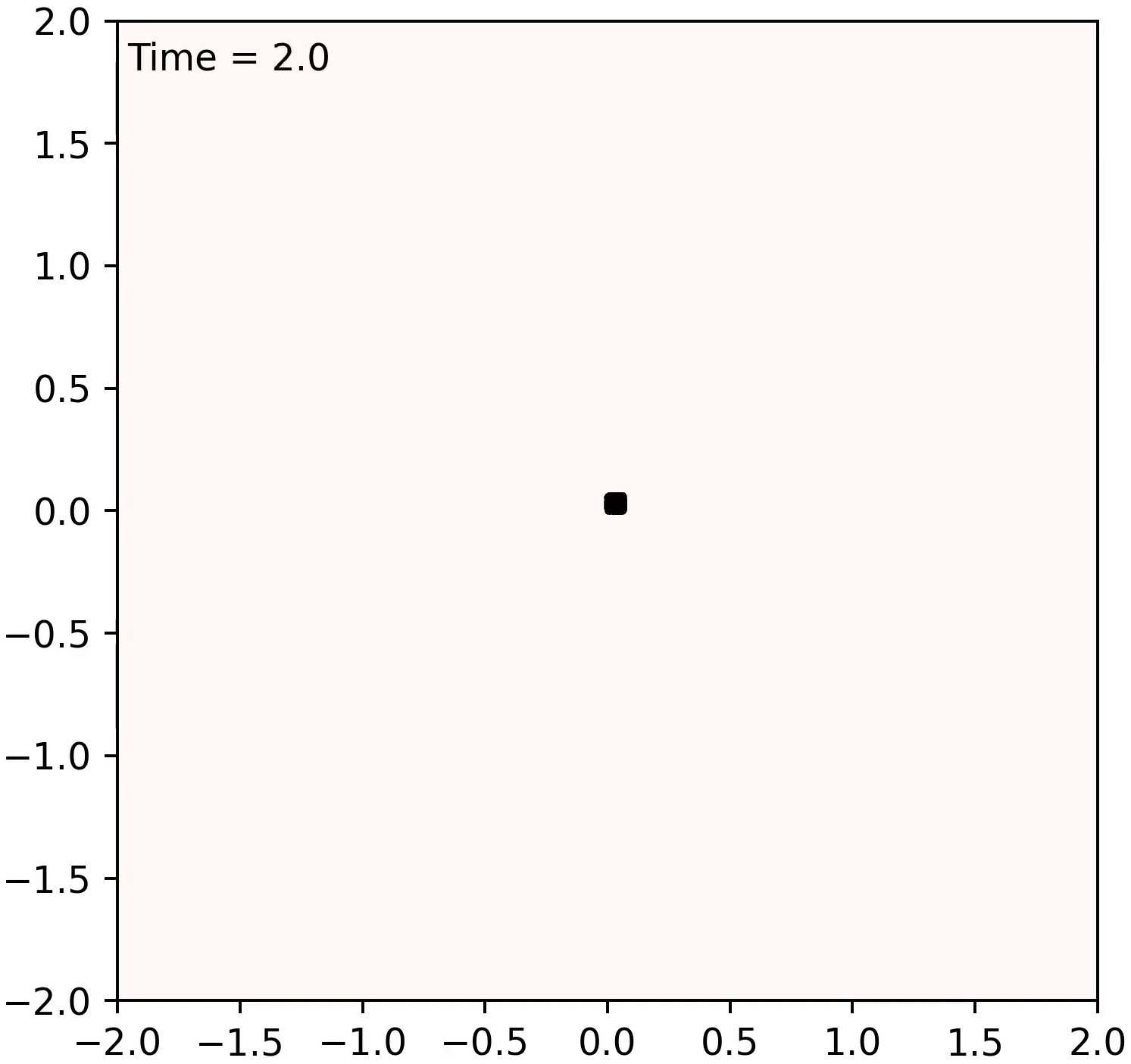}
     \end{subfigure}
     \begin{subfigure}[t]{0.32\columnwidth}\hfil
         \centering
         \includegraphics[width=\linewidth]{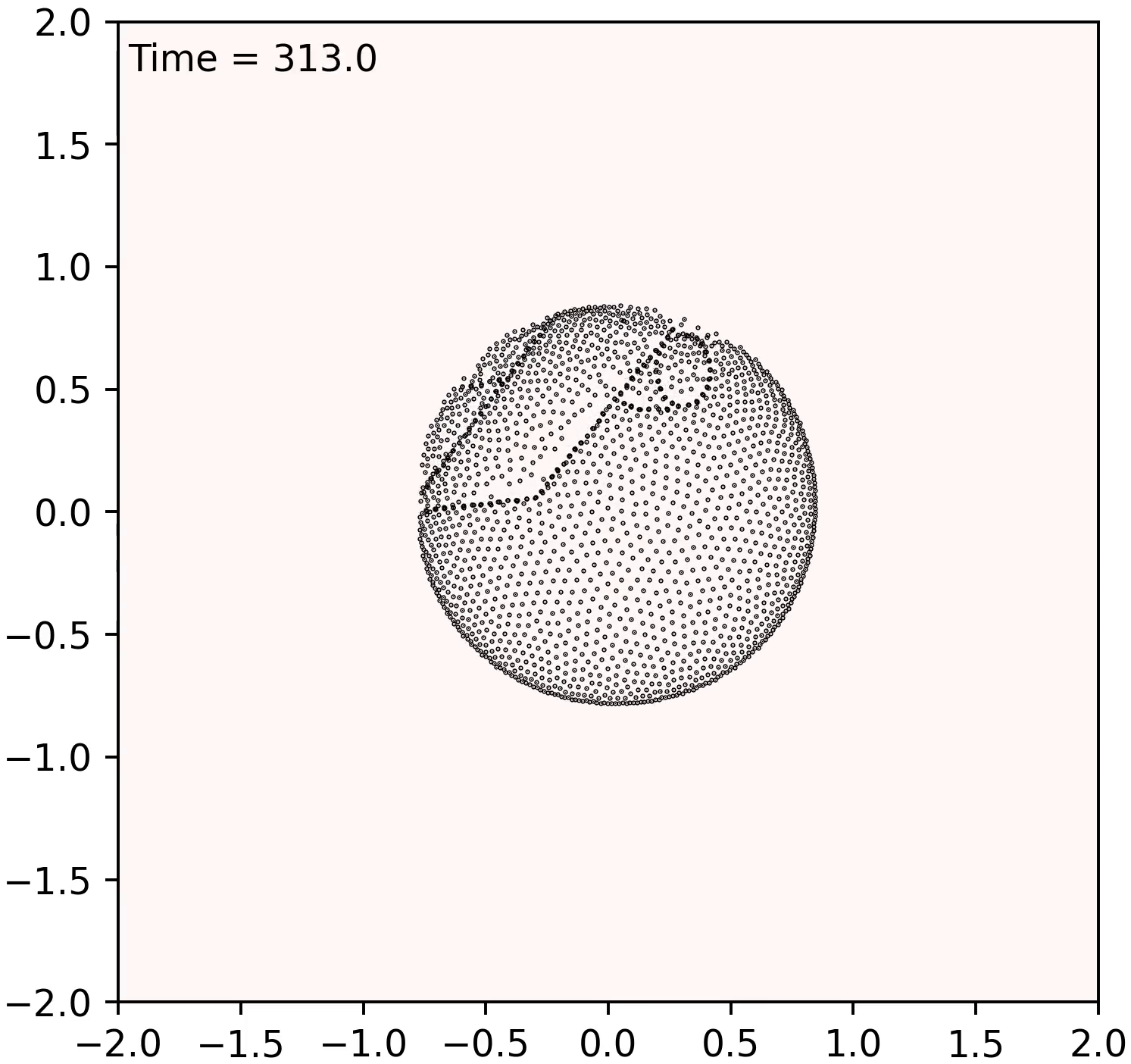}
     \end{subfigure}
     \begin{subfigure}[t]{0.32\columnwidth}\hfil
         \centering
         \includegraphics[width=\linewidth]{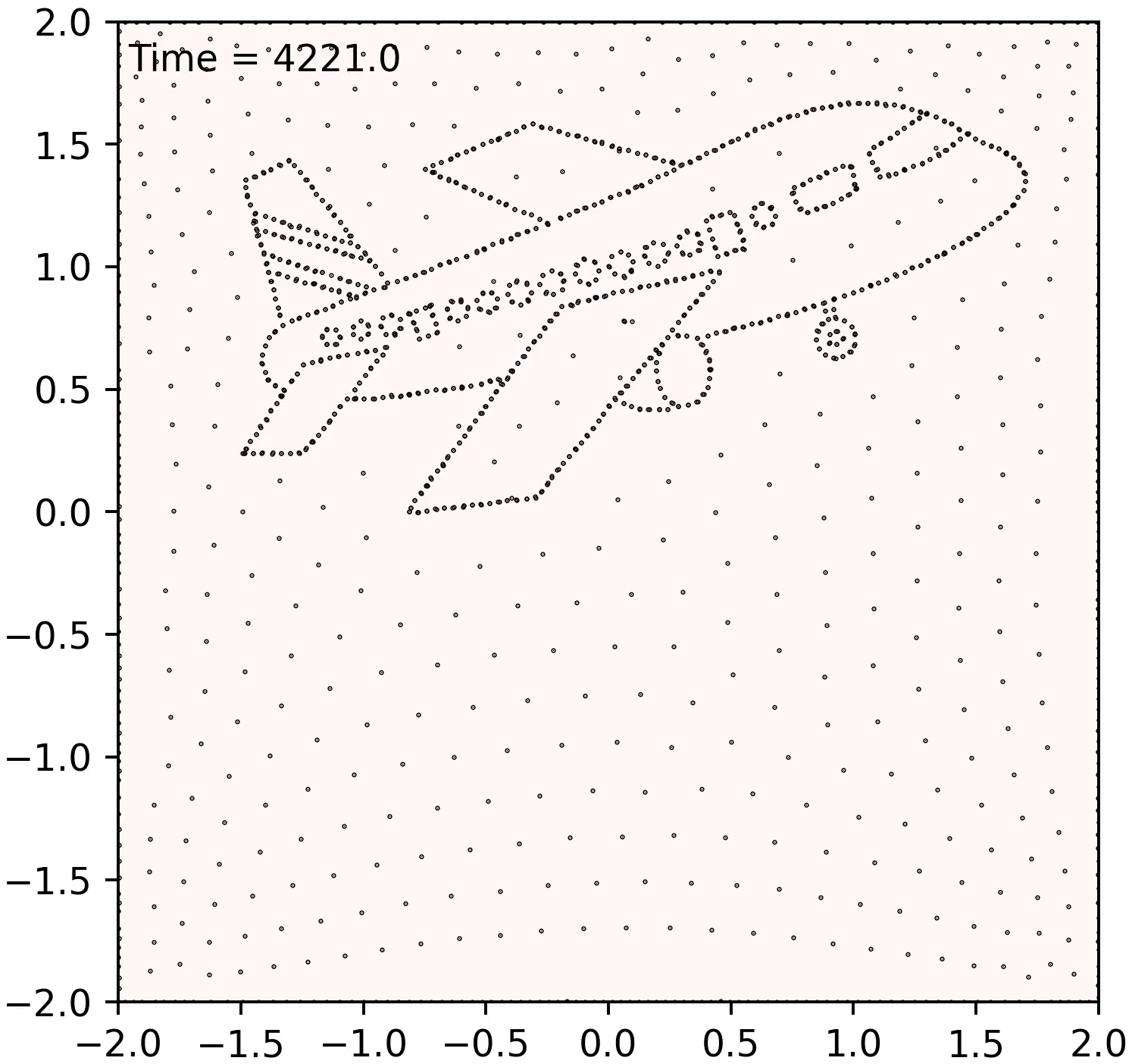}
     \end{subfigure}
     \end{adjustwidth}
    \caption{Limited sensing range electrostatic target assignment with many targets and agents, demonstrating a high degree of precision and scalability. The total demand of the targets is 1240, and the number of agents is 2000. Targets not drawn for the sake of visual clarity. Agent sensing radius, not depicted in the image, is 1/10th of the bounding box.}
     \label{fig:Sim3_electrostatic_target_assignment_plane_figure}
\end{figure}

Figures \ref{fig:Sim2_electrostatic_target_assignment} and  \ref{fig:Sim3_electrostatic_target_assignment_plane_figure}  illustrate a target assignment task using electrostatic attraction-repulsion dynamics. %In general, we found that these dynamics complete target assignment faster, and require fewer agents than the signal coverage-based target assignment approach. Signal coverage, however, solves a more general class of task allocation problems. Furthermore, we only tested a finite number of signal functions $f$, and so the performance difference might be due to our choice of signal function. We leave these questions for future work.

\subsubsection{Scalar field coverage}
\label{section:Fieldattractionrepulsion}

\begin{figure}[!ht]
    \begin{adjustwidth}{\figureWidthAdjustment}{\figureWidthAdjustment}
     \centering
     \begin{subfigure}[t]{0.32\columnwidth}\hfil
         \centering
         \includegraphics[width=\linewidth]{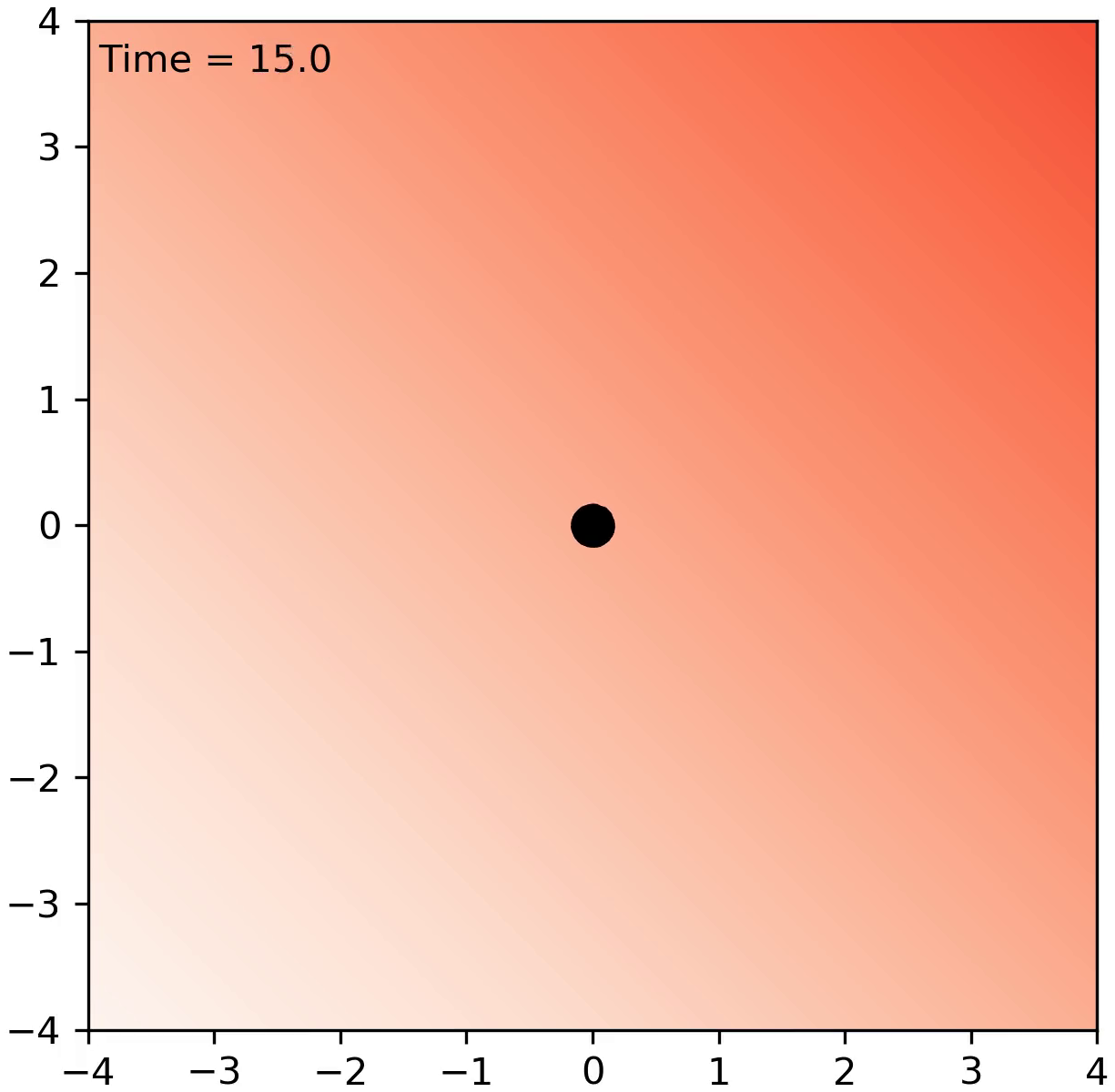}
     \end{subfigure}
 %    \begin{subfigure}[t]{0.32\columnwidth}\hfil
  %       \centering
  %       \includegraphics[width=\linewidth]{Figures/Sim_10/10_A3.png}
  %   \end{subfigure}
     \begin{subfigure}[t]{0.32\columnwidth}\hfil
         \centering
         \includegraphics[width=\linewidth]{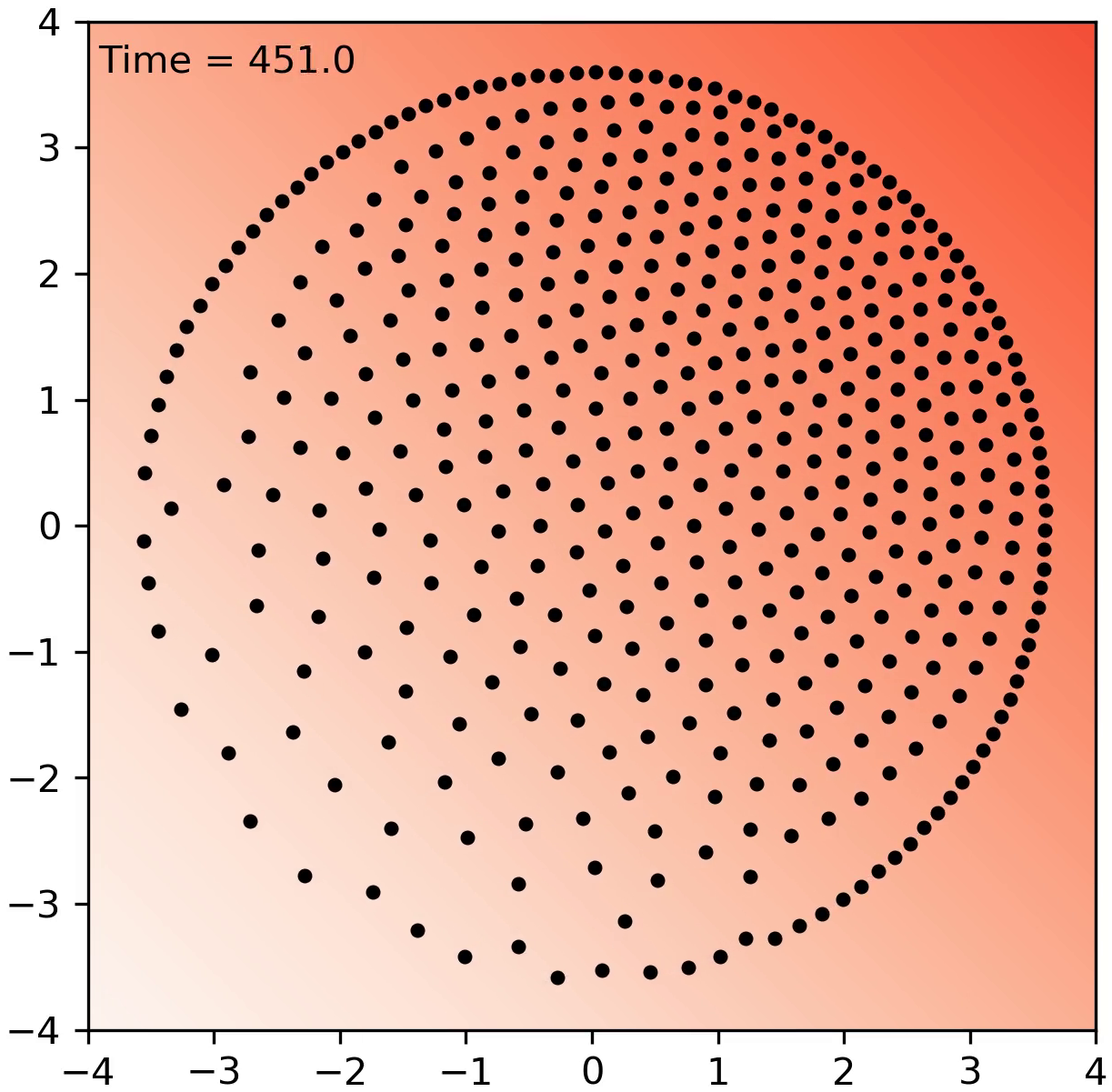}
     \end{subfigure}
  %   \begin{subfigure}[t]{0.32\columnwidth}\hfil
   %      \centering
   %      \includegraphics[width=\linewidth]{Figures/Sim_10/10_B3.png}
   %  \end{subfigure}
   %  \begin{subfigure}[t]{0.32\columnwidth}\hfil
   %      \centering
  %       \includegraphics[width=\linewidth]{Figures/Sim_10/10_C2.png}
   %  \end{subfigure}
     \begin{subfigure}[t]{0.32\columnwidth}
         \centering
         \includegraphics[width=\linewidth]{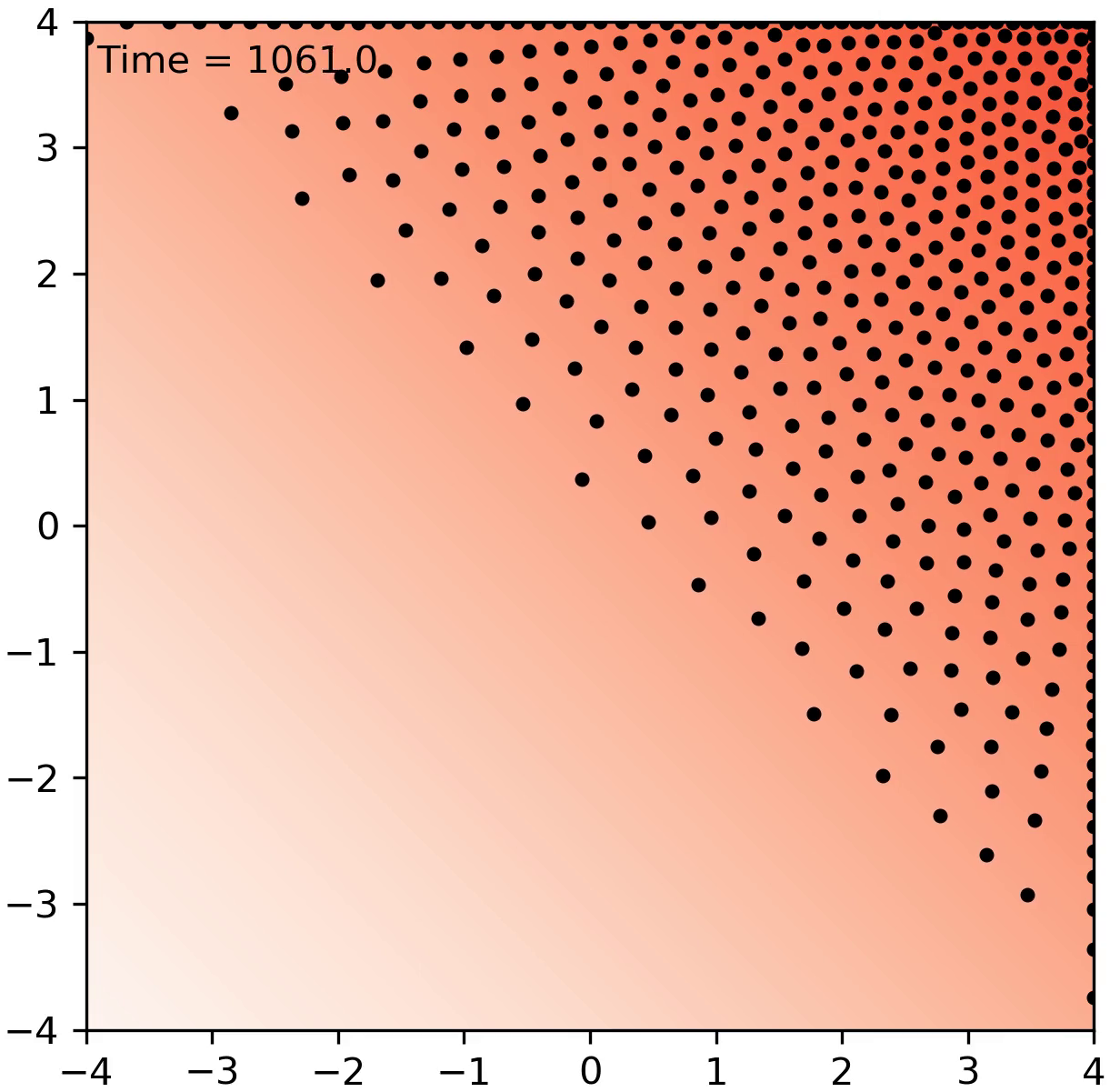}
     \end{subfigure}
     
    \begin{subfigure}[t]{0.32\columnwidth}\hfil
         \centering
         \includegraphics[width=\linewidth]{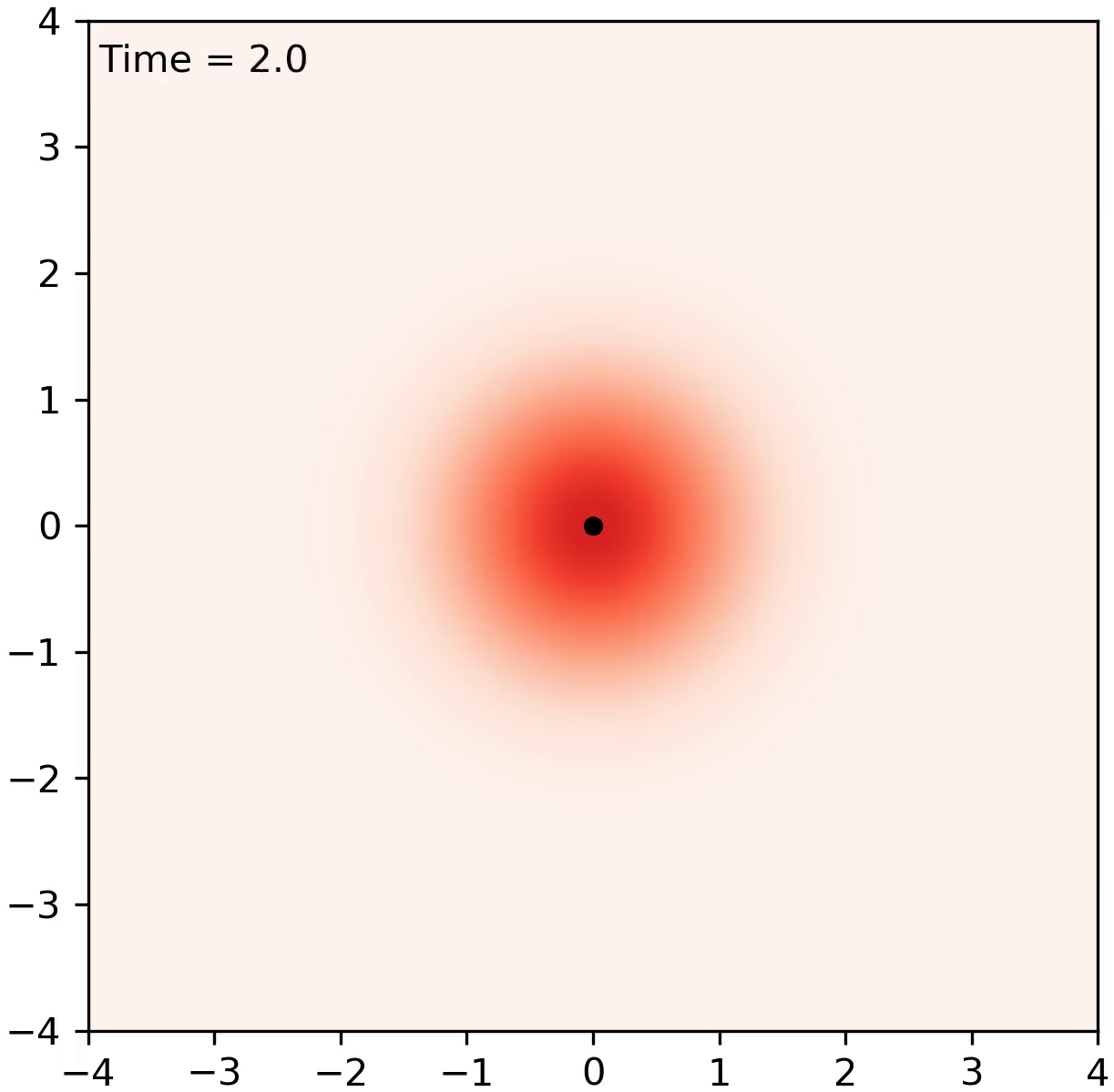}
     \end{subfigure}
   %  \begin{subfigure}[t]{0.32\columnwidth}\hfil
    %     \centering
    %     \includegraphics[width=\linewidth]{Figures/Sim_11/11_A3.png}
   %  \end{subfigure}
   %  \begin{subfigure}[t]{0.32\columnwidth}\hfil
    %     \centering
    %     \includegraphics[width=\linewidth]{Figures/Sim_11/11_B1.png}
   %  \end{subfigure}
     \begin{subfigure}[t]{0.32\columnwidth}\hfil
         \centering
         \includegraphics[width=\linewidth]{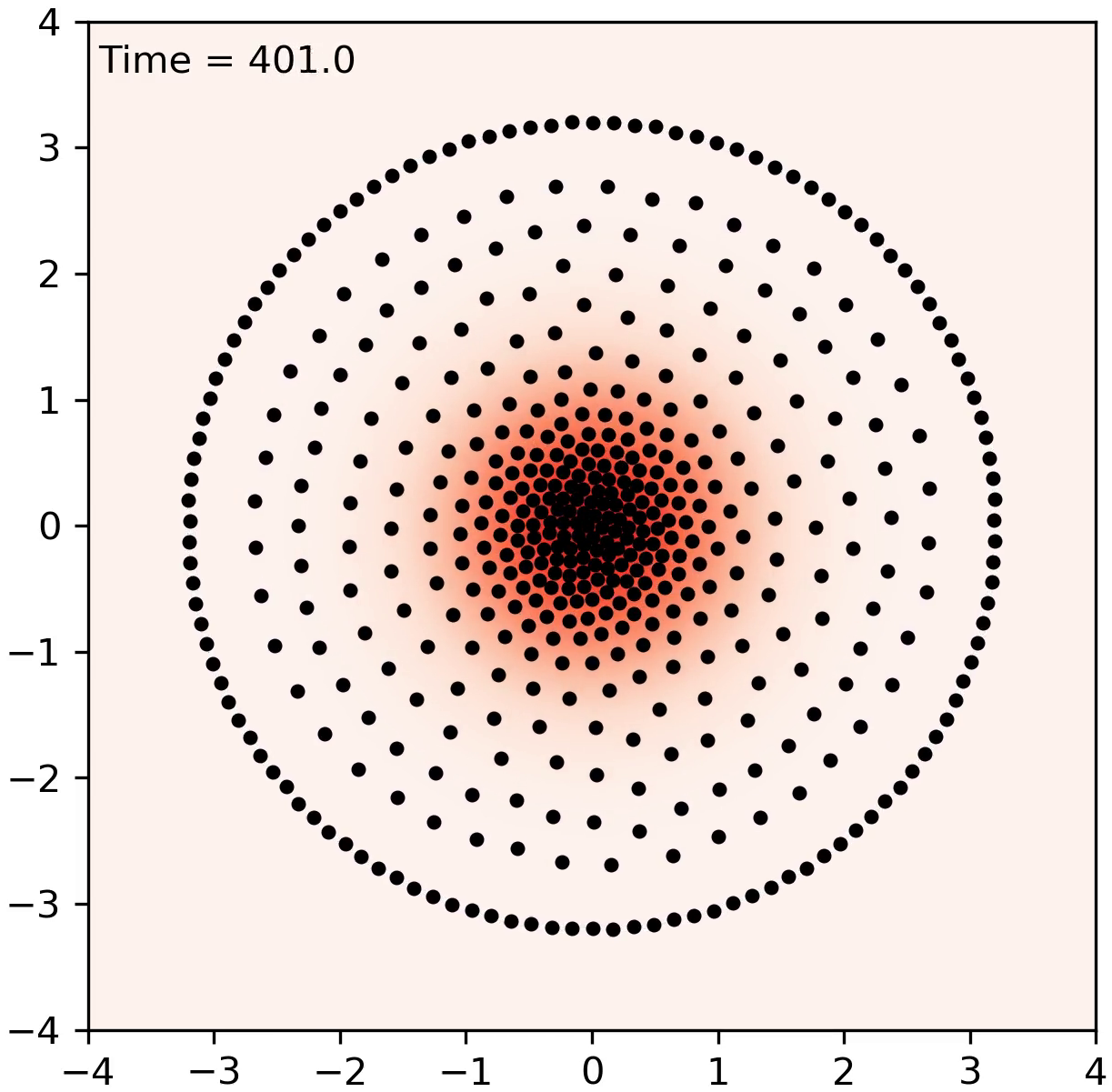}
     \end{subfigure}
  %   \begin{subfigure}[t]{0.32\columnwidth}\hfil
   %      \centering
  %       \includegraphics[width=\linewidth]{Figures/Sim_11/11_C2.png}
  %   \end{subfigure}
     \begin{subfigure}[t]{0.32\columnwidth}
         \centering
         \includegraphics[width=\linewidth]{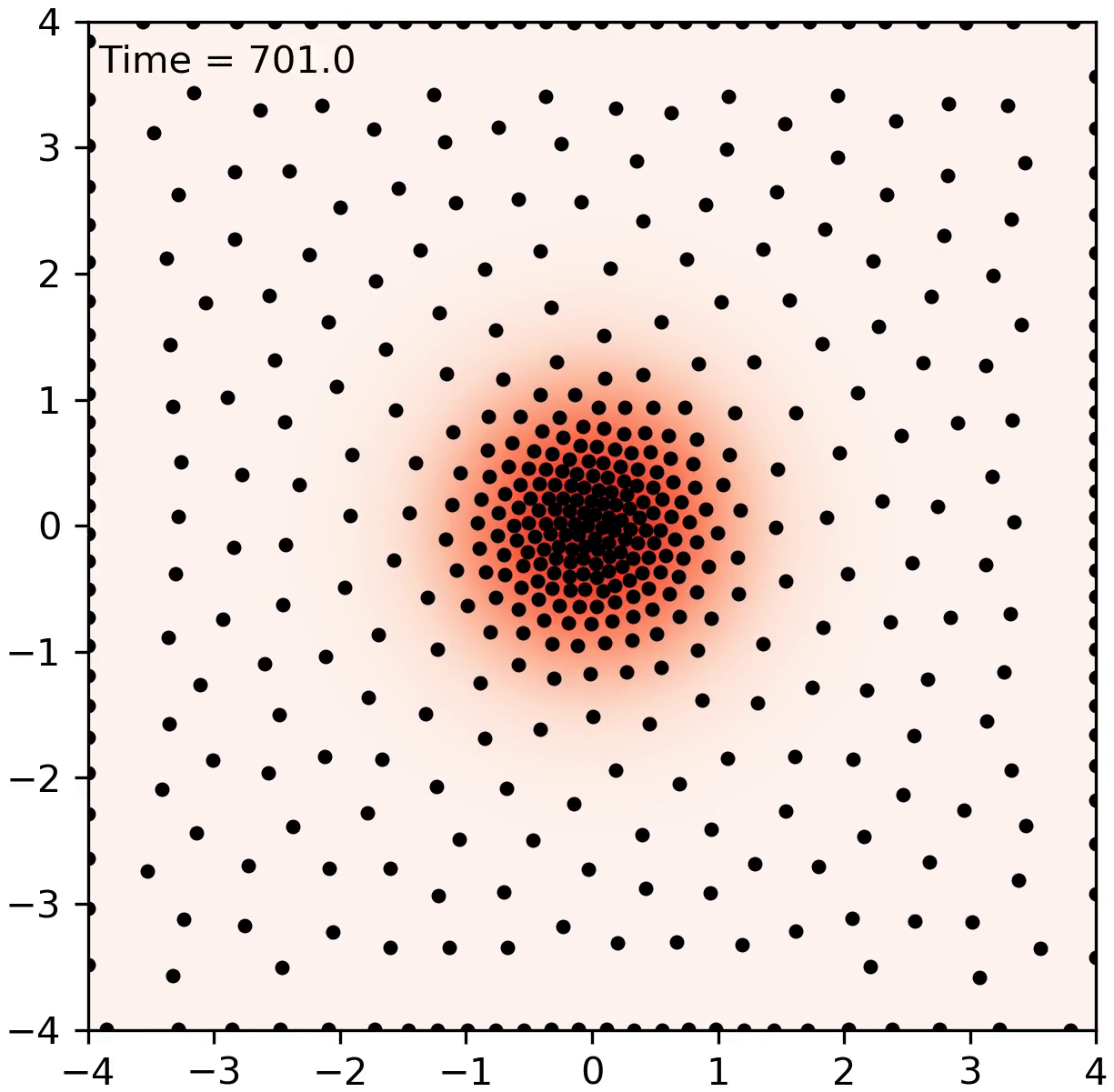}
    \end{subfigure}
     \end{adjustwidth}
     \caption{500 agents covering a linear demand function of the form $\Phi(x,y) = c (x+y)$ (top row), or an exponential demand function of the form $\Phi(x,y) = c \cdot e^{-\Lambda (x^2+y^2)}$ (bottom row). The heat map depicts $\Phi(x,y)$.}
     \label{fig:Sim10_Spatial_demand_linear}
\end{figure}

Moving away from target assignment, let us consider the problem of covering an arbitrary scalar field with agents (Figure \ref{fig:Sim10_Spatial_demand_linear}). One way to do this is to let $\Phi(x,y)$ be our scalar field and model the repulsion forces of the agents the same as any of our proposed approaches to target assignment. When using Section \ref{section:electrostatictargetassignment}'s electrostatic AR-dynamics as the base model we get the following dynamics for the $i$th agent:

\begin{equation}
\begin{split}
        \frac{\partial \mathcal{G}}{\partial \vec{p}_i} (\vec{\textbf{q}})& =  \sum_{1  \leq j \leq \mathcal{N}}  \frac{\overrightarrow{p_i p_j}}{\lVert \vec{p}_i - \vec{p}_j \rVert^2} - \frac{\partial }{\partial \vec{p}_i}\Phi(\vec{p_i})
\end{split}
\label{eq:scalarfielddynamicsinfinitevisibility}
\end{equation}

For example, when $\Phi(x,y) = ax+by$ is a linear scalar field we get that $\frac{\partial }{\partial \vec{p}_i}\Phi(\vec{p_i}) = \begin{bmatrix}
           a \\
           b 
         \end{bmatrix}$ and so agents will move in the direction $\begin{bmatrix}
           a \\
           b 
         \end{bmatrix}$ while spreading out due to repulsion forces. When $\Phi(x,y) = c \cdot e^{-\Lambda (x^2+y^2)}$ the agents will gather near the maximum of $\Phi(x,y)$ in a diffusive fashion (note that we also used exponential demand functions in signal coverage, but because the error function $\Psi$ is different in signal coverage, we get different dynamics).

%%%%%%%%%%%%%%%%%%%%%%%%%%%%%%%
%\begin{figure}[h]
   % \begin{adjustwidth}{\figureWidthAdjustment}{\figureWidthAdjustment}
   %  \centering
   %  \begin{subfigure}[t]{0.32\columnwidth}\hfil
    %     \centering
    %     \includegraphics[width=\linewidth]{Figures/Sim_11/11_A1.png}
  %   \end{subfigure}
   %  \begin{subfigure}[t]{0.32\columnwidth}\hfil
    %     \centering
    %     \includegraphics[width=\linewidth]{Figures/Sim_11/11_A3.png}
   %  \end{subfigure}
   %  \begin{subfigure}[t]{0.32\columnwidth}\hfil
    %     \centering
    %     \includegraphics[width=\linewidth]{Figures/Sim_11/11_B1.png}
   %  \end{subfigure}
   %  \begin{subfigure}[t]{0.32\columnwidth}\hfil
   %      \centering
   %      \includegraphics[width=\linewidth]{Figures/Sim_11/11_B3.png}
  %   \end{subfigure}
  %   \begin{subfigure}[t]{0.32\columnwidth}\hfil
   %      \centering
  %       \includegraphics[width=\linewidth]{Figures/Sim_11/11_C2.png}
  %   \end{subfigure}
  %   \begin{subfigure}[t]{0.32\columnwidth}
  %       \centering
 %        \includegraphics[width=\linewidth]{Figures/Sim_11/11_C3.png}
  %   \end{subfigure}
 %    \end{adjustwidth}
%     \caption{500 agents covering an exponential demand function of the form $\Phi(x,y) = c \cdot e^{-\Lambda (x^2+y^2)}$.}
%     \label{fig:Sim11_Spatial_demand_exp}
%\end{figure}

%%%%%%%%%%%%%%%%%%%%%%%%%%%%%%%%%%%%%%%%%%%%%%%%%%%%%%%%%%%%%%%%%%%%%%%%
\section{Discussion}

The goal of this work was a preliminary exploration of the topic of geometric task allocation for swarms of oblivious, decentralized agents with limited sensing range. We discussed two task allocation problems: signal coverage, wherein robots must ``cover'' some a priori unknown demand profile $\Phi(x,y)$ with signals they emit, and target assignment, wherein targets (representing, e.g., search and rescue tasks) are placed in discrete, a priori unknown locations, and agents must find and the targets and move to their location. 

Our solutions to these problems use attraction-repulsion dynamics: agents repulse each other, encouraging exploration of the environment, and tasks attract agents, causing the agents to organize according to the tasks' requirements. In the case of signal coverage, we showed such dynamics are naturally derived from gradient descent over the squared error $\Psi(x,y,\vec{\textbf{q}}) = \big(\Phi(x,y) - \sum_{\substack{i = 1}}^{\mathcal{N}} \tilde{d} (x-x_i,y-y_i) \big)^2$. Surprisingly, we could not find any work in the literature obtaining attraction-repulsion dynamics from the squared difference of signal functions, and in future work, we are interested in further investigating the mathematical theory underlying this result. We proposed two different approaches to target assignment. The first is treating it as a signal coverage problem with highly concentrated environmental signals at the location of each target. The second is electrostatic target assignment via Coulomb's law. Finally, we related our solutions to a general form of attraction-repulsion dynamics, and discussed related applications such as scalar field coverage.

%In future work, we are interested in finding more task allocation problems to which the attraction-repulsion approach can be applied. We are further interested in developing the mathematical theory behind this approach, specifically asking in what situations we can prove performance guarantees.

\bibliography{bib}
\bibliographystyle{plain}

\end{document}